\documentclass[preprint2]{aastex}

\slugcomment{Published in The Astrophysical Journal 697 (2009), 807-823}

\begin{document}

\title{Grain Alignment in OMC1 as 
Deduced from Observed Large Circular Polarization 
\footnote{This is an author-created, un-copyedited version of 
an article accepted for publication 
in The Astrophysical Journal 697 (2009), pp.807-823. 
IOP Publishing Ltd is not responsible for any errors or omissions 
in this version of the manuscript or any version derived from it. 
The definitive publisher authenticated version is available online 
at doi:10.1088/0004-637X/697/1/807.  
\ \ \ \ \ \ \ \ \ \ \ \ \ \ \ \ \ \ \ \ \ \ \ \ \ \ \ \ \ \ \ \ \ \ 
\copyright 2009. The American Astronomical Society.}} 

\author{M. Matsumura}
\affil{Faculty of Education, Kagawa University, Takamatsu, 
    Kagawa 760-8522, Japan}
\email{matsu@ed.kagawa-u.ac.jp}

\and

\author{P. Bastien}
\affil{D\'epartement de physique \&
    Centre de recherche en astrophysique du Qu\'ebec,
    Universit\'e de Montr\'eal, C.P.6128, Succursale Centre-ville, 
    Montr\'eal, Qu\'ebec, H3C 3J7, Canada}
% \email{aastex-help@aas.org}

\begin{abstract}
The properties of polarization in scattered light by aligned ellipsoidal 
grains are investigated with the Fredholm integral equation method (FIM) and 
the T-matrix method (Tmat), 
and the results are applied to the observed circular polarization in OMC1.  
We assume that the grains are composed of silicates and and ellipsoidal 
(oblate, prolate, or tri-axial ellipsoid) in shape with a typical axial 
ratio of 2:1.  
The angular dependence of circular polarization $p_{c}$ on directions of 
incident and scattered light is investigated 
with spherical harmonics and associated Legendre polynomials.
The degree of circular polarization $p_{c}$ also depends on the Rayleigh 
reduction factor $R$ which is a measure of imperfect alignment.
We find that $p_{c}$ is approximately proportional to $R$ 
for grains with $|m|x_{eq} \la 3 - 5$, where $x_{eq}$ is 
the dimensionless size parameter and $m$ is the refractive index of the grain.
Models that include those grains can explain the observed large 
circular polarization in the near-infrared, $\approx 15\%$, 
in the south-east region of the BN object (SEBN) in OMC1, 
if the directions of incidence and scattering of light is optimal, and 
if grain alignment is strong, i.e., $R \ga 0.5$.
Such a strong alignment cannot be explained by the Davis-Greenstein 
mechanism;  we prefer instead an alternative mechanism driven by 
radiative torques.
If the grains are mixed with silicates and ice, 
the degree of circular polarization $p_{c}$ decreases 
in the 3 $\mu$m ice feature, 
while that of linear polarization increases.
This wavelength dependence is different from
that predicted in a process of dichroic extinction.

\end{abstract}


%% subject header.  Objects should be in the appropriate "individual"
%% headers (e.g. quasars: individual, stars: individual, etc.) with the
%% additional provision that the total number of headers, including each
%% individual object, not exceed six.  The \objectname{} macro, and its
%% alias \object{}, is used to mark each object.  The macro takes the object
%% name as its primary argument.  This name will appear in the paper
%% and serve as the link's anchor in the electronic edition if the name
%% is recognized by the data centers.  The macro also takes an optional
%% argument in parentheses in cases where the data center identification
%% differs from what is to be printed in the paper.

 --- polarization --- scattering

% \keywords{globular clusters: general ---
% globular clusters: individual(\objectname{NGC 6397},
% \object{NGC 6624}, \objectname[M 15]{NGC 7078},
% \object[Cl 1938-341]{Terzan 8})}


\section{Introduction}

Polarimetry of young stellar objects (YSOs) such as 
T Tau and Herbig Ae/Be stars and their circumstellar material  
provides information about the distribution of matter and/or 
the configuration of magnetic field in their environment, confirming that 
polarimetry is an important tool to study the physical processes 
in star forming regions.
Large linear polarization up to 20\% has been observed 
in many YSOs at visible and infrared wavelengths
\citep[see][for a review]{tam05}. 
In linear polarization maps of YSOs, the polarization vectors 
near the star are often aligned in a direction parallel to the disk  
while they show a circular pattern far from the star.
These patterns are usually interpreted 
as due to single and/or multiple scattering by circumstellar grains 
\citep{bas96}. 

Circular polarimetry of YSOs has also been carried out,  
although fewer measurements are available than for linear polarimetry. 
Aperture circular polarimetry shows that circular polarization 
in several YSOs is small, i.e., the degree of circular polarization $p_c$ 
is $\approx0.01-0.1\%$ in the optical \citep{nadeau86,menard88,bas89}.
Mapping observations of the Chamaeleon infrared nebula \citep{gled96} and 
the GSS 30 reflection nebula \citep{chrys97} show 
circular polarization of 1 or 2 \% in the near IR.
More recently, much larger circular polarization has been found in near IR, 
i.e., $p_c \approx 5\%$ in the $H$-band in R CrA \citep{clark00},
$15\%$ in the $K$-band in the south-east region of BN  
(hereafter SEBN) in OMC1 \citep{bailey98,chrys00,busch05}, 
and  $23\%$ in the $K$-band in NGC6334 \citep{menard00}. 

At least four possible mechanisms 
based on light scattering/extinction by dust grains 
have been proposed for explaining circular polarization: 
\begin{enumerate}
\item Multiple scattering by non-aligned (spherical or nonspherical) grains,
\item dichroic extinction by aligned nonspherical grains,
\item single scattering by aligned nonspherical grains, and 
\item multiple scattering by aligned nonspherical grains. 
\end{enumerate}
Multiple scattering models by non-aligned grains 
(Mechanism 1) predict a circular polarization $p_c$ of
at most a few percent 
\citep[][and references therein]{bas96},
which is comparable to observed values 
in the Chamaeleon infrared nebula \citep{gled96} and
in the GSS 30 reflection nebula \citep{chrys97}.
Therefore, one may deduce that circumstellar grains in 
those objects are not aligned, 
although low circular polarization will be expected 
even if grains are aligned, depending on various conditions. 
However, larger circular polarization as observed in R CrA 
\citep{clark00}, OMC1 \citep{bailey98,chrys00,busch05},  
and NGC6334 \citep{menard00}  
cannot be explained by models without grain alignment. 

Dichroic polarization (Mechanism 2) occurs when light is transmitted 
through a medium where grains are aligned in a given direction. 
If the direction of alignment does not change along the line of sight, 
only linear polarization is produced  
while circular polarization arises when the direction of 
alignment changes \citep{martin78}.
The linear and circular 'interstellar polarization' 
observed when stellar light passes through diffuse clouds 
is explained by dichroic extinction.
\citet{lucas05} showed that strong circular polarization can occur 
with extinction, if grains are small dielectric particles.
\citet{busch05} found a correlation between $J-K$ color 
and circular polarization $p_c$ in the $K$-band in OMC1 (their Fig.7) 
which led them to favor dichroic extinction 
as the mechanism for producing circular polarization in this object.
However, \citet{min91} suggested that major part of the $J$ flux is 
due to scattered light from the Trapezium stars and free-free radiation. 
An inspection of 2MASS images shows that the $J$ flux varies by only  
$\approx$ 0.3 mag in the region of their Fig.7 in \citet{busch05}, 
while the $K$ flux varies by $\approx$ 1 mag. 
Since the effect of extinction should be larger in $J$ than in $K$, 
the $J-K$ color variation does not seem to be due to extinction. 
The correlation between $J-K$ and $p_c$ can be interpreted such that 
the brighter part in the $K$-band, i.e., the region dominated 
by scattered light, is more polarized circularly.  
We therefore prefer the other mechanisms, i.e. Mechanism 3 or 4.

Circular polarization can also be produced by single scattering 
by aligned grains (Mechanism 3) \citep{sch73,martin78,dolgi78,dolgi92,ms96a, 
bailey98,chrys00,gled00,mb04,bm05}.
If the grain is very elongated or flattened  
and if the imaginary part of the refractive index is moderately large,
circular polarization will be large, e.g. $>30\%$, 
even for grains smaller than wavelength, i.e., in the 
Rayleigh approximation (see Fig.5 in \cite{gled00}).
When the size of dielectric grains is relatively large, 
circular polarization becomes large, even for grains 
which are not much elongated or flattened \citep{gled00}.
Those grains show different angular dependence of circular polarization
from that for the Rayleigh approximation \citep{gled00}. 
\citet{chrys00} explained the ratio of 
linear to circular polarization, or the ellipticity of polarization, 
of SEBN using silicate or organic refractory grains 
with sizes of $0.1-1.0 \mu$m.
They ruled out metallic grains because those particles 
do not explain the observed wavelength dependence of $p_c$.

Calculations of multiple scattering by aligned grains 
(Mechanism 4) have been carried out recently \citep{wolf99,whit02,lucas03}.
The optical depth from the YSOs to the observer is usually much 
larger than unity, so multiple scattering should occur.
Observations of thermal emission from grains in dense regions 
show significant linear polarization in the submm implying that those 
grains are nonspherical and aligned \citep{hilde95}. 
Therefore the study of multiple scattering (in the visible and 
near IR) by aligned grains should be very rewarding.
However these models are much more complex than those based on 
the previous three mechanisms.
The models have now at least three additional parameters, 
two angles for the direction of alignment and the degree of alignment,
compared to models for Mechanism 1 (multiple scattering 
by non-aligned grains). 

The mechanisms proposed until now to explain the circular polarization 
in SEBN are dichroic extinction (Mechanism 2) 
and single scattering by aligned nonspherical grains (Mechanism 3). 
We know by now that multiple scattering is required for explaining 
observations in most YSO environments \citep[e.g.,][]{BfM88,BfM90}
and therefore our ultimate goal is to study Mechanism 4. 
As a step in this direction, we explore in this paper Mechanism 3. 
Our results will be useful, among other things, for comparing results 
between single and multiple scattering, i.e., Mechanisms 3 and 4. 
We use two methods, the Fredholm integral equation method (FIM) 
\citep{holt78,ms91,ms96a,ms96b} and 
T-matrix method (Tmat) \citep{mish00,mht00}.
FIM can be applied to tri-axial ellipsoidal particles, 
while Tmat is very efficient in evaluating the scattering properties 
of spheroidal particles.
We first compare the results of FIM with those of Tmat 
and show that the two numerical methods give essentially the same results 
under the same conditions (Section~\ref{fim_tmat}). 
Since circular polarization in the presence of weak alignment has not been 
investigated extensively so far, except for analytical studies by 
\citet{dolgi78} and \citet{dolgi92}, 
we present models for aligned grains (Section~\ref{align-models}).  
In our previous papers \citep{mb04,bm05}, we showed the dependence of 
$p_c$ on the scattering angle, 
i.e. the angle between the directions of incidence and scattering.
Here we use spherical harmonics and associated Legendre polynomials, 
and study further the angular dependence of circular polarization 
not only on the scattering angle but also on the directions of 
incident and scattered beams (Section~\ref{expa}). 
We compare our results to SEBN polarization data and show that
the observed linear and circular polarization in the $K$ and $L$-bands 
can be explained if the grains are composed of silicates with 
radii of 0.15-1.5 $\mu$m, an axial ratio of 2:1, 
and the Rayleigh reduction factor $R \ga 0.5$ (Section~\ref{parameters}). 
We discuss the wavelength dependence of polarization 
(Section~\ref{wavelength-dep}), 
the grain shape and its degree of alignment (Section~\ref{degree}), 
and the direction of alignment (Section~\ref{direction}).
As a step toward Mechanism 4, we comment on the scattering properties 
if the incident beam is already polarized (Section~\ref{pol-inc}), 
and also on dichroic polarization (Section~\ref{dich-pol}).
Finally, we assume that the grains are a mixture of silicates and ice 
and examine a possible polarization variation in the 3 $\mu$m 
ice band feature (Section~\ref{band-feature}).

\section{Calculations} \label{cal}

\subsection{FIM and Tmat}  \label{fim_tmat}
The Fredholm integral equation method (FIM) is one of the solutions 
to the light scattering problem, and is applicable to homogeneous and 
isotropic tri-axial ellipsoidal particles with an axial ratio of a few 
\citep{holt78,ms91,ms96a,ms96b}.
The light scattering process is expressed as 
a Fredholm-type integral equation with a singular kernel. 
\citet{holt78} removed the singularity by using a Fourier transform 
which leads to a linear equation that can be solved numerically.
This method is known to be numerically stable \citep{holt78}.
The major part of the FIM calculation is independent of 
the directions of incidence and scattering, 
thus FIM is efficient for scattering calculations 
in many different directions. 
We have developed a numerical code using FIM; with our most recent version (version 2.1) 
on a desktop computer we can calculate scattering functions 
for a size parameter $x_{\rm max} (= 2\pi a_{\rm max}/\pi)$ up to $\approx 10$, 
where $a_{\rm max}$ is the largest axis of the ellipsoid.

A popular solution of the light scattering problem 
for axisymmetrical particles is 
the T-matrix (Tmat) method \citep{mish00,mht00}. 
Tmat expands the incident and scattered waves 
with the vector spherical wave functions, and 
their coefficients are related by a matrix called a "T-matrix". 
One can evaluate the T-matrix and then solve the scattering  
problem numerically (Chap.6 in \citet{mht00}).
Although Tmat can be applied to particles of any shape,
the formulation is simpler for axisymmetrical particles  
and the public domain codes are restricted to such particles.
This is the most efficient method and it can be applied to particles 
with a size parameter up to $\approx 100$.
In our simulations we used the Fortran program "ampld.new.f" dated 
04/03/2003 written by Mishchenko.   %%%  (\url{http://www.giss.nasa.gov/~crmim}). 

Both FIM and Tmat are rigorous solutions and give
essentially the same results. 
As an example, Fig.~\ref{f_vs_t} shows the degree of  
circular polarization $p_c$ (see eq.(\ref{p_c}) below) 
for an oblate particle with an axial ratio of 2:1 and $m=1.7$. 
The results from the two methods are in very good agreement. 

The geometrical configuration adopted in this paper is 
shown in Fig.~\ref{sph1}. 
The grain is at the origin of the coordinate system and 
the direction of incident light $I$ is defined by the angle $\theta_i$ 
with respect to the symmetry axis.
The scattered light goes in direction $S$ defined by $(\theta_s,\phi_s)$. 
The scattering angle $\Theta_{sca}$, which is the angle between $I$ and $S$,  
can be calculated by solving the spherical triangle $AIS$.

The transformation or the Mueller matrix of 
the Stokes parameters is written as
\begin{equation}
\left( \begin{array}{cccc}
	I_s &  Q_s & U_s & V_s
	\end{array} \right)^T
= F_{jk} \cdot 
\left( \begin{array}{cccc}
	I_i &  Q_i & U_i & V_i
	\end{array} \right)^T
\label{stks}
\end{equation}
where $j=1,...,4$, $k=1,...,4$, and the suffixes $i$ and $s$ stand for 
'incident' and 'scattered', respectively. 
The elements $F_{jk}$ are calculated with FIM or Tmat 
(see Appendix A for the sign of circular polarization).

In the following calculations for oblate grains 
we first make a table of $F_{jk}$, either with FIM or Tmat, 
and then we evaluate the values of $F_{jk}$ in arbitrary directions 
with a Spline interpolation \citep{press92}.
The table of $F_{jk}$ contains data of $9\times17\times17 (=2601)$ points 
in the 3-D parameter space of $(\cos \theta_i, \cos \theta_s, \phi_s)$
for $\theta_i=0-90\arcdeg$, $\theta_s=0-180\arcdeg$, and 
$\phi_s=0-180\arcdeg$, respectively;  
this corresponds to sampling in $\approx10\arcdeg$ spacing. 
This angular resolution seems to be sufficient to obtain $\la$ 1\% accuracy 
on the polarization of scattered light from our grain models.
For prolate grains, we assume that the grain spins around its 
short axis and make a table of $F_{jk}$ for
$(\cos \theta_i, \cos \theta_s, \phi_s)$. 
For tri-axial ellipsoidal grains, we first evaluate $F_{jk}$ with FIM 
not only for $(\cos \theta_i, \cos \theta_s, \phi_s)$, 
but also for the azimuthal angle $\phi_i$ of the incident direction. 
We then integrate $F_{jk}$ for $\phi_i$, 
assuming that the grain spins around its shortest axis 
and provide a table for $(\cos \theta_i, \cos \theta_s, \phi_s)$. 
The rest of the process is the same as that for oblate grains.

We have used the following symmetry relations in our calculations to 
reduce the CPU time significantly.
The values of $F_{31}, F_{41}, F_{32}, F_{42}, F_{13}, F_{23}, 
F_{14}$, and $F_{24}$ change their sign while the others remain the same 
if the angles $(\theta_i,\theta_s,\phi_s)$ are replaced by 
$(180\arcdeg-\theta_i,180\arcdeg-\theta_s,\phi_s)$  or by 
$(\theta_i,\theta_s,360\arcdeg-\phi_s)$.
It should be noted that the element $F_{41}$ becomes zero 
if the angles $(\theta_i,\theta_s,\phi_s)$ take specific values
in the axisymmetrical models (Table 1).  
This implies that circular polarization can change 
sign many times according to the geometrical configuration.

\subsection{Imperfect Alignment} \label{align-models}

\subsubsection{Specific Cone Angle Model (S-model)} \label{smodel} 

Grain alignment would not be perfect in interstellar/circumstellar space;  
it may be perfect or imperfect, depending on alignment mechanism and 
various factors \citep[e.g.][for alignment by radiative torques]{LH07,Hoang08}.
To explore the effect of imperfect alignment, we consider two models. 
The first one is the "specific cone angle model" (S-model) 
and the second one is the "continuous distribution of cone angle model 
(CD-model, discussed in Sect.~\ref{cd} below).
In the S-model, we assume that 
(1) the directions of the spin axis, the angular momentum,
and the maximum moment of inertia coincide (direction $A$ 
in Fig.~\ref{sph2}), and 
(2) the direction $A$ is distributed around another direction, 
the direction $B$ in Fig.~\ref{sph2}, 
while keeping the polar angle $\theta_{a}'$ constant. 
According to assumption (1) 
the symmetry axis is parallel to $A$ for oblate grains and 
perpendicular to it for prolate grains. 
For tri-axial ellipsoidal grains, the shortest axis coincides with $A$.
%% compared to other alignment timescales \citep{DM76,laz00}.
Conventionally, the Rayleigh reduction factor $R$ is used to express 
the degree of alignment and is related to the angle $\theta_{a}'$ as 
\begin{equation}
R = (3{ \cos }^2\theta_{a}' - 1)/2 = 1 - (3/2){ \sin }^2\theta_{a}'.
\label{defR}
\end{equation}
The amount of dichroic polarization is exactly proportional to $R$ 
in the Rayleigh approximation and is approximately proportional 
to $R$ in interstellar grain models. 
Perfect alignment corresponds to $R=1$ and non-alignment to $R=0$.

Fig.~\ref{sph2} shows the relation between the various angles 
in our scattering model: 
the direction of incidence $I$ is defined by the angle $\theta_i'$, 
and that of scattering $S$ by the angles $\theta_s'$ and $\phi_s'$  
with respect to the direction of alignment $B$. 
If there is no other scattering after this one, 
$S$ corresponds to the direction toward the observer. 
The direction $A$ is defined by $\theta_a'$ and $\phi_a'$. 
The matrix $F_{jk}$ is a function of  $\theta_i, \theta_s$, 
and $\phi_s$, and these angles are calculated with trigonometry 
from $\theta_i'$, $\theta_s'$, $\phi_s'$, $\theta_a'$ and $\phi_a'$.
In the S-model, we numerically integrate the values $F_{11},...,F_{44}$ 
over the range $\phi_a'=0$ to $180\arcdeg$.   

For natural incident light, i.e., $I_i \ne 0, Q_i=U_i=V_i=0$, 
it is sufficient to consider the first column of the matrix 
$F_{k1} (k=1,...,4)$.  
The degree of linear polarization $p_l$, the position angle $PA$, 
and the degree of circular polarization $p_c$ are calculated with
\begin{equation}
p_l = \sqrt{F_{21}^2 + F_{31}^2} / F_{11}
\end{equation}
\begin{equation}
PA  = 90\arcdeg - \arctan (F_{21}/F_{31})/2 - \alpha_2
%%%%%%% PA  = \pi/2 - \arctan (F_{21}/F_{31})/2 - \alpha_2
\end{equation}
\begin{equation}
p_c = F_{41} / F_{11},
\label{p_c}
\end{equation}
where $\alpha_2$ is the azimuthal angle $BSA$ in Fig.\ref{sph2} 
and the position angle $PA$ is defined in the usual way, 
increasing counterclockwise as seen from the observer $S$ 
with respect to the reference direction $B$. 
We discuss the scattering properties for polarized incident light 
in Section~\ref{pol-inc}. 

Circular polarization by precessing spheroidal grains has been 
calculated by \citet{gled00}, so we compare their results with ours 
in Fig.~\ref{gm}. 
The agreement between the two results for prolate grains is satisfactory 
since the difference between the two calculations is less than 
$\approx0.1$\%.
However, results for oblate grains show a small but significant difference 
of $\approx1$\% whose cause is unknown. 

\citet{gled00} argued that the circular polarization $p_c$ produced 
by prolate grains is much smaller than that by oblate grains 
if they are imperfectly aligned. 
They first searched for the direction of the maximum circular 
polarization in "perfect" alignment, i.e., the long axis is directed 
in a direction.  
Next they calculated $p_c$ in imperfect alignment 
assuming the direction of maximum $p_c$ is the same as 
that for perfect alignment. 
If the grains are relatively small,
the angular dependence of $p_c$ does not change and 
$p_c$ is proportional to the Rayleigh reduction factor $R$ 
in imperfect alignment, as shown in Section~\ref{precessing}. 
However, the grain size in their model (their LG model) is 
larger, and the dimensionless size parameter reaches 6 
for the maximum size in their size distribution. 
Therefore, the angular dependence of $p_c$ 
varies as the degree of alignment changes.
The prolate grain in the "perfect" alignment takes 
its maximum $p_c$ value  
at $(\theta_i',\theta_s',\phi_s')=(50.0\arcdeg,45.6\arcdeg,145.3\arcdeg)$, 
while in the "perfectly spinning" alignment the maximum occurs 
at $(\theta_i',\theta_s',\phi_s')=(75.5\arcdeg,82.8\arcdeg,112.5\arcdeg)$. 
The former case corresponds to the results of \citet{gled00}   
and is presented by the open circles and the dotted line in Fig.~\ref{gm}. 
The latter is shown by the broken line in Fig.~\ref{gm}  
and is much larger than the former when $\theta_a' < 30\arcdeg$.
Although prolate grains show lower $p_c$ than oblate grains, 
the difference is not as large as shown by \citet{gled00} 
under optimum condition.

\subsubsection{Continuous Distribution of Cone Angle Model (CD-model)} \label{cd}

The S-model is simple and provides a good first approximation 
but is not sufficient to describe polarization in weak alignment.
In the S-model, we set $R=0$, or $\theta_a'=54.7$\arcdeg, for nonalignment 
and expect to find $p_c=0$. However, the calculated polarization is 
not zero but typically a few percent in our models. 
This leads us to introduce the second model in which 
the direction $A$ is distributed not only 
in $\phi_a'$, but also in $\theta_a'$.  
We assume that the cone angle $\theta_a'$ is distributed homogeneously 
from 0 to $\theta_{a0}'$. 
Nonalignment corresponds to $\theta_{a0}'=90\arcdeg$ and perfect alignment 
appears when $\theta_{a0}'=0\arcdeg$. 
It should be noted that such modeling is not new and has been used by 
\citet{whit02} previously for models with $\theta_{a0}'=10\arcdeg$ and $30\arcdeg$. 
In the CD-model, the Rayleigh reduction factor $R$ is related 
with $\theta_{a0}'$ as 
\begin{equation}
 R = ( \cos^2 \theta_{a0}' + \cos \theta_{a0}' ) / 2.
\end{equation}

\subsection{Expansion with Spherical Harmonics \&  Associated Legendre Polynomials} 
 \label{expa}
We examine the dependence of $F_{41}$ or $p_c (=F_{41}/F_{11})$ 
on the angles $\theta_i, \theta_s$, and $\phi_s$ 
by using spherical harmonics and associated Legendre polynomials
for perfectly aligned grains (Section~\ref{formulation}) and 
precessing grains (Section~\ref{precessing}).

\subsubsection{Formulation} \label{formulation}
If the angle $\theta_i$ is constant, then the quantities $F_{41}$ and 
$p_c$ are functions of $\theta_s$ and $\phi_s$ only and can be written  
with spherical harmonics $Y_n^m(\theta_s,\phi_s)$. 
For $F_{41}$ we write
\begin{equation}
F_{41}(\theta_i,\theta_s,\phi_s) = 
	\sum_{n=1}^{\infty} \sum_{m=1}^{n} b_{nm} Y_n^m(\theta_s,\phi_s), 
\label{F41}
\end{equation}
where the coefficients $b_{nm}$ are calculated as
\begin{equation}
  b_{nm} = \int_{\phi_s=0}^{2\pi}\int_{\theta_s=0}^{\pi}F_{41}(\theta_i,\theta_s,\phi_s) Y_n^m(\theta_s,\phi_s)d\theta_{s}d\phi_s.
\end{equation}
Since $F_{41}$ is an odd function in $\phi_s$, only odd components for $\phi_s$  
or the terms in sin$(m\phi_s)$ are necessary: 
\begin{eqnarray}
\lefteqn{ Y_n^m(\theta_s,\phi_s)= } \nonumber \\
& & 	\sqrt{\frac{(n-m)!(2n+1)}{2\pi(n+m)!}} P_n^m(\cos\theta_s)\sin(m\phi_s). \nonumber \\
\end{eqnarray}
The dependence on the angle $\theta_i$ is carried by the coefficients $b_{nm}$. 
We thus expand $b_{nm}$ with associated Legendre polynomials 
$P_l^k(\cos\theta_i)$  where $k$ is zero or a positive integer. 
If we choose $k=1$, the first term of the expansion for $F_{41}$, 
or the term with $(n,m,l)=(1,1,2)$, 
coincides with the expression for the Rayleigh approximation.
We therefore set $k=1$, and write $b_{nm}$ as
\begin{equation}
  b_{nm}(\theta_i) =
	\sum_{l=1}^{\infty}a_{nml} \sqrt{\frac{2l+1}{2l(l+1)}} P_l^1(\cos\theta_i),
\end{equation}
where 
\begin{equation}
  a_{nml} = \sqrt{\frac{2l+1}{2l(l+1)}} \int_0^{\pi}b_{nm}(\theta_i) P_l^1(\cos\theta_i)d\theta_i.
\end{equation}
Combining eqs.(7)-(11) yields 
\begin{equation}
  F_{41}(\theta_i,\theta_s,\phi_s) = 
	\sum_{l=1}^{\infty} \sum_{n=1}^{\infty}  \sum_{m=1}^{n} 
	a_{nml}f^{nml}(\theta_s,\phi_s,\theta_i),
\end{equation}
where 
\begin{eqnarray}
	a_{nml} =  \int_{\theta_i=0}^{\pi} \int_{\phi_s=0}^{2\pi} \int_{\theta_s=0}^{\pi}
    	F_{41}(\theta_i,\theta_s,\phi_s) \nonumber \\
	\times f^{nml}(\theta_{s},\phi_s,\theta_i)d\theta_s d\phi_s d\theta_i,
\end{eqnarray}
% \begin{equation}
%   a_{nml} = 
% 	\int_{\theta_i=0}^{\pi} \int_{\phi_s=0}^{2\pi} \int_{\theta_s=0}^{\pi}
% 	F_{41}(\theta_i,\theta_s,\phi_s)f^{nml}(\theta_{s},\phi_s,\theta_i) 
% 	d\theta_s d\phi_s d\theta_i,
% \end{equation}
and 
\begin{eqnarray}
f^{nml}(\theta_s,\phi_s,\theta_i) = 
        \sqrt{\frac{(2l+1)(n-m)!(2n+1)}{4\pi l(l+1)(n+m)!}}  \nonumber \\
	\times P_l^1(\cos\theta_i) P_n^m(\cos\theta_s)\sin(m\phi_s).
\label{f^nml}
\end{eqnarray}
%\begin{equation}
%  f^{nml}(\theta_s,\phi_s,\theta_i) = 
%        \sqrt{\frac{(2l+1)(n-m)!(2n+1)}{4\pi l(l+1)(n+m)!}} 
%	P_l^1(\cos\theta_i)
%	P_n^m(\cos\theta_s)\sin(m\phi_s).
%\label{f^nml}
%\end{equation}
The functions $f^{nml}$ as given by eq.(\ref{f^nml}) are orthonormal, 
\begin{eqnarray}
\lefteqn{
	\int_{\theta_i=0}^{\pi} \int_{\phi_s=0}^{2\pi} \int_{\theta_s=0}^{\pi}
	f^{nml}(\theta_i,\theta_s,\phi_s) }  \nonumber \\
& &	\times f^{n'm'l'}(\theta_i,\theta_{s},\phi_s) d\theta_s d\phi_s d\theta_i 
        = \delta_{nn'}\delta_{mm'}\delta_{ll'}, \nonumber \\
\end{eqnarray}
%\begin{equation}
%	\int_{\theta_i=0}^{\pi} \int_{\phi_s=0}^{2\pi} \int_{\theta_s=0}^{\pi}
%	f^{nml}(\theta_i,\theta_s,\phi_s)f^{n'm'l'}(\theta_i,\theta_{s},\phi_s) 
%	d\theta_s d\phi_s d\theta_i 
%        = \delta_{nn'}\delta_{mm'}\delta_{ll'},
%\end{equation}
where $\delta_{nn'}$ is Kronecker's $\delta$-function.

With this relation 
\begin{eqnarray}
\lefteqn{P_n^m(-\cos\theta_s)P_l^1(-\cos\theta_i) = }  \nonumber \\
&& -(-1)^{n+m+l}P_n^m(\cos\theta_s)P_l^1(\cos\theta_i),
\end{eqnarray}
% \begin{equation}
% P_n^m(-\cos\theta_s)P_l^1(-\cos\theta_i) =
% -(-1)^{n+m+l}P_n^m(\cos\theta_s)P_l^1(\cos\theta_i),
% \end{equation}
we see that the product $P_n^m(\cos\theta_s)P_l^1(\cos\theta_i)$ in eq.(\ref{f^nml}) 
changes sign when $n+m+l$ is even but not when it is odd. 
The quantities $p_c$ and $F_{41}$ show the same properties to the case of
$n+m+l$ even, i.e., they change sign 
when $\theta_i$ is replaced by $\pi-\theta_i$ and 
$\theta_s$ by $\pi-\theta_s$, while $\phi_s$ unchanged.
Therefore the coefficients $a_{nml}$ vanish when $n+m+l$ is odd,  
and $p_c$ and $F_{41}$ can be written with only the even terms in $n+m+l$.
As a result we can write the first few terms of $F_{41}$ as
\begin{eqnarray}
\lefteqn{F_{41}(\theta_i,\theta_s,\phi_s) = }  \nonumber \\
  & & \frac{3}{4}\sqrt{\frac{5}{\pi}}a_{112}\sin\theta_i\cos\theta_i\sin\theta_s\sin\phi_s +  \nonumber \\
  & & \frac{3}{4}\sqrt{\frac{5}{\pi}}a_{211}\sin\theta_i\sin\theta_s\cos\theta_s\sin\phi_s +  \nonumber \\
  & & \frac{15}{8}\sqrt{\frac{1}{\pi}}a_{222}\sin\theta_i\cos\theta_i\sin^2\theta_s\sin2\phi_s +\cdots. \nonumber \\
\label{F41-2}
\end{eqnarray}
When the Rayleigh approximation is valid, only the first term ($a_{112}$)
is necessary for the expression of $F_{41}$. 

The degree of circular polarization $p_c$ can be expressed the same way 
as explained above and many examples are shown in Fig.~\ref{anml1}.
For dielectric particles, the overall dependence of $a_{nml}$ on $|m|x_{eq}$ 
is similar, except for two cases: prolate grains computed with 
the "perfect" alignment (Fig.~\ref{anml1}d) and oblate grains 
with axial ratio of 3:1 (Fig.~\ref{anml1}g). 
The term in $a_{112}$ is the dominant one for $|m|x_{eq} \la 2$, 
which indicates that the Rayleigh approximation holds. 
Around $|m|x_{eq} \approx 2.6$, the sign of $a_{112}$ changes, 
and for $|m|x_{eq} \ga 2.6$, both the terms in $a_{211}$ and $a_{112}$ are 
large. Since $|m|x_{eq}$ is a size parameter normalized by the wavelength 
inside the particles, the properties of circular polarization are determined by 
the transmitted wave within the particles. 
Absolute values of $a_{112}$ and $a_{211}$ are largest 
when $|m|x_{eq} \approx 3$ or 4 and they start to decrease 
for larger values of $x_{eq}$, 
for the cases we computed for dielectric particles with $m=1.7$ or $1.7+0.1i$.

The values of $a_{nml}$ for metallic particles are functions of 
$x_{eq}$, and not of $mx_{eq}$, 
i.e.  the overall dependence of their $a_{nml}$ on $x_{eq}$ 
is similar (Fig.~\ref{anml1}h and i). 
Therefore, the mechanism to produce circular polarization 
in metallic particles seems different 
from that in the dielectric ones.
The absolute values of $a_{112}$ and $a_{211}$  are largest 
for $x_{eq} \approx 1$ 
and they decrease when $x_{eq} \ga 1$ or $x_{eq} \approx 0$. 

\subsubsection{Precessing Grains}  \label{precessing}
When the grain is incompletely aligned, 
the values of $F_{41}$ and $p_c$ decrease, and 
the angular dependencies change.
For precessing grains (S-model, Section~\ref{smodel}), 
we change the Rayleigh reduction factor $R$, 
and investigate the variation of the function 
$f^{nml}(\theta_s',\phi_s',\theta_i')$  (eq.(\ref{f^nml})).
We write the modified functions as $g^{nml}(\theta_i',\theta_s',\phi_s',R)$ 
under the effect of precession, which is identical to $f^{nml}$ for $R=1$.
We expand $g^{nml}$ with spherical harmonics 
and associated Legendre polynomials already included in $f^{nml}$, 
in the same way as explained in Section~\ref{formulation}:
\begin{eqnarray}
 \lefteqn{g^{nml}(\theta_i',\theta_s',\phi_s',R) =  }  \nonumber \\
 & &	\sum_{l'=1}^{\infty} \sum_{n'=1}^{\infty}  \sum_{m'=1}^{n} 
 	a_{n'm'l'}^{nml}(R)f^{n'm'l'}(\theta_s',\phi_s',\theta_i'). \nonumber \\
\end{eqnarray}
For $(n,m,l)=(1,1,2)$ or $(2,1,1)$, we find that $g^{nml}$ can be written as 
\begin{equation}
  g^{nml}(\theta_i',\theta_s',\phi_s',R) 
  = R\cdot f^{nml}(\theta_i',\theta_s',\phi_s'). 
\label{R}
\end{equation}
The angular dependence in these cases does not change and 
the amplitude is proportional to $R$ under incomplete alignment.

Terms with higher $(n,m,l)$ are required  
for the expansion of $g^{nml}$ for higher values of $(n,m,l)$, 
i.e. $(n,m,l) \ne (1,1,2)$ nor $(2,1,1)$. 
The examples of $a_{n'm'l'}^{nml}(R)$ shown in Fig.~\ref{anml2}  
have been numerically evaluated 
for $(n,m,l)=(2,2,2)$, $(2,1,3)$, $(3,1,2)$, $(3,2,1$), $(3,2,3)$, 
and $(3,3,2)$.
As expected, the value of $a_{nml}^{nml}$ is unity for 
perfect alignment ($R=1$).
The values of $a_{n'm'l'}^{nml}$ for $(n',m',l') \ne (n,m,l)$ are 
not large especially when $R$ is small. This means that large circular 
polarization cannot be expected from large grains if alignment is poor.  
 
\subsection{Results for Imperfectly Aligned Grains}
To evaluate how the results change when grains are not perfectly aligned, 
we assume the S-model (Section~\ref{smodel}) and 
calculate the means of $F_{41}(R) / F_{41}(1)$ 
(Figs.~\ref{cmp17-13}a and c) and the standard deviations 
(Figs.~\ref{cmp17-13}b and d) 
for values of $R=0.25, 0.5$ or $0.75$ for oblate grains 
and one case for prolate grains.  
We assume that the symmetry axis of prolate grains is perpendicular
to spin axis, and calculate the ratios of values $F_{41}$
for those grains over $F_{41}$ in perfect alignment.
This condition is expressed as $R=-0.5$ in Fig.~\ref{cmp17-13}.
We took 1785 cases or less into account out of 2601 
(see Section~\ref{fim_tmat}) 
because we excluded 816 cases where $F_{41}(1)=0$ due to
symmetry (see Table 1) and also other cases where $F_{41}(1) \approx 0$.
In the Rayleigh approximation or when $mx_{eq} \la 2$, 
the means of $F_{41}(R) / F_{41}(1)$ are nearly equal to $R$ 
and the standard deviations are relatively small, 
as expected (Section~\ref{precessing}).
For larger $mx_{eq}$, the means become smaller, and 
the standard deviations larger. 
However, the means for $m=1.7$ are nearly equal to $R$ up to 
$mx_{eq} \approx 4$, except for a drop near $mx_{eq} \approx 2.6$ 
(Fig.~\ref{cmp17-13}a). 
The value of $F_{41}$ for $m=1.7$ can be expressed well 
by using only the component of $a_{112}$ and $a_{211}$ (Fig.\ref{anml1}a).  
Since these components are proportional to $R$, 
$F_{41}$ is also proportional to $R$. 
At $mx_{eq} \approx 2.6$, this proportionality fails 
because $a_{112}$ is zero and the relative contributions 
from other terms become large.
The maximum circular polarization $|p_c|$ is large when $mx_{eq} \ga 3$ 
while the degree of linear polarization $p_l$ is small 
(Figs.\ref{cmp17-13}e and f).

For the dielectric particles considered here, 
one or two peak(s) is (are) expected for $p_c$ 
as a function of the angles $(\theta_i, \theta_s, \phi_s)$,  
because the values of $a_{112}$ and $a_{211}$ are 
much larger than higher order terms, as shown above.
It is highly possible that one of the peaks in the $p_c$ can explain 
the high circular polarization observed in SEBN of OMC1.
We thus investigate the directions which correspond to the maximum value 
of $p_c$ as a function of $mx_{eq}$ for particles with an axial ratio of 2:1, 
in a parameter space of 
$(\theta_i, \theta_s, \phi_s) = (0-180\arcdeg, 0-180\arcdeg, 0-180\arcdeg)$  
(Fig.~\ref{angles}). 
The values of angles for minimum $p_c$, 
i.e. the negative of maximum value, can be retrieved by using 
a symmetry relation (Section~\ref{fim_tmat}).
In the Rayleigh approximation, 
the circular polarization $p_c$ is expected to reach its maximum value 
when $(\theta_i, \theta_s, \phi_s) = (135\arcdeg, 90\arcdeg, 90\arcdeg)$
(see eq.(B1) in Appendix B), 
while Figs.~\ref{angles}b-d show $(129\arcdeg, 90\arcdeg, 90\arcdeg)$ 
when $mx_{eq} \approx 0$.
The difference in the values of $\theta_i$ is due to the coarse sampling 
in the angle.
One can see systematic variations of the angles corresponding to 
maximum $p_c$ value as a function of $mx_{eq}$,    
although those variations are not clear for larger $mx_{eq}$.
For $mx_{eq} \approx 3-5$, the value of $p_c$ takes its maximum in
$(\theta_i, \theta_s, \phi_s) \approx (40-60\arcdeg, 40-90\arcdeg, 100-140\arcdeg)$, 
and this property does not depend on shape considered here.

\section{Discussion: the SEBN Region in OMC1} 
We now compare our results to the observed large polarization in SEBN of OMC1. 
We use the circular polarimetry data by \citet{chrys00} 
(their Table 2), and the linear polarimetry by \citet{min91} (their Table 4).  
These tabulated values are observed in the box region of size 
$3\arcsec\times3\arcsec$ centered at $22\arcsec$ east 
and $6\arcsec$ south from the BN object. 

\subsection{Grain Parameters}  \label{parameters}
A few papers have studied the parameters of grains which can explain 
the observed polarization in the SEBN region so far. 
\citet{chrys00} showed that oblate grains of silicate with an axial ratio of 2:1 
can explain the ratio of circular to linear polarization observed in SEBN 
provided the lower cut-off and the upper cut-off in a power law size distribution
are set to $a_{min}=0.1\mu$m and $a_{max}=1.0\mu$m, 
where the size is the equivalent radius of a sphere with the same volume. 
\cite{lucas05} found that a distribution of silicate grains with sizes up to 
$a_{max} = 0.75\mu$m could produce circular polarization in the $K$-band 
if their axial ratio is larger than 3:1. 
However, reproducing the model by \citet{chrys00} or that by \citet{lucas05}, 
we find that the degree of circular polarization $p_c$ is too small to explain 
the observed one ($p_c=9.4$\%) in the $L$-band ($\lambda=3.6\mu$m), 
though the circular polarization $p_c=15\%$ in the $K$-band can be explained.
The angular dependence of those grains varies so greatly 
from the $K$ to the $L$-band that it is impossible to explain  
the observed circular polarization in the $K$ and $L$-bands simultaneously 
if the angles of incidence and scattering do not depend on wavelength.

We thus examine the effects of grain size parameters 
in a power law size distribution and have found 
that the upper cut-off $a_{max}$ is the most effective for changing 
the circular polarization $p_c$ in the $L$-band, 
while $a_{min}$ and the power index are less effective. 
Fig.~\ref{size} shows the circular polarization $p_c$ of 
oblate grains with axial ratios of 2:1 or 3:1 
as a function of the upper cut-off $a_{max}$, 
setting $a_{min}$ as $a_{min} = 0.1a_{max}$, 
and the power index as $-3.5$. 
If the axial ratio is 2:1 (solid lines), the observed values of 
polarization ($p_l=38$\% and $p_c=15$\%) in the $K$-band can be 
explained if $a_{max} > 1.0\mu$m, 
while those in the $L$-band ($p_l=57$\% and $p_c=9.4$\%)
can be explained only if $a_{max} \ga 1.5\mu$m.  
The grains with the axial ratio 3:1 (dashed lines) can explain 
the observed circular polarization in the $L$-band 
even if $a_{max} \approx 1\mu$m. 
However, the calculated linear polarization is higher than 
the observations in both the $K$ and $L$-bands.   
Therefore, we set $a_{max} = 1.5\mu$m, $a_{min} = 0.15\mu$m, and 
the axial ratio is 2:1 as a typical grain model 
in the following calculations. 

As for the refractive index, 
we use the data for "smoothed astronomical silicate" \citep{WD01,D03}.
Although the refractive index $m$ of the silicate depends little 
on wavelength $\lambda$ in $0.5 \la \lambda \la 5\mu$m, 
eg. $m=1.661+0.035i$ in the $K$-band ($\lambda=2.2\mu$m) and 
$m=1.638+0.041i$ in the $L$-band ($\lambda=3.6\mu$m), 
the "effective" refractive index shows large variation around $\lambda \approx 3 \mu$m  
if ice is mixed in. We investigate this effect in Section~\ref{band-feature}.

\subsection{Wavelength Dependence}  \label{wavelength-dep}
We show the wavelength dependence of $p_l$, $PA$, $p_c$, 
and $p_c/p_l$ in Fig.~\ref{wv-dep2}, for models of oblate, prolate, 
and ellipsoidal grains composed of silicate with an axial ratio of 
2:1 (2:$\sqrt{2}$:1 for ellipsoids).
Here we use the CD-model (Section~\ref{cd}) for grain alignment, 
and we take the Rayleigh reduction factor $R$ as 0.5 or 1.0. 
We assume optimum directions of incidence and scattering 
to obtain the largest circular polarization.
The overall trends of the observed quantities seem to be explained.  
However, the observed linear polarization $p_l$ is systematically larger 
than the computed values in most of our models. 
If we decrease the size, the fit for $p_l$ becomes better, 
but the circular polarization $p_c$ is smaller than the observed values.
The agreement for position angle $PA$ is not good (Fig.~\ref{wv-dep2}b)  
and the reason is not clear. 
The values of both $p_l$ and $p_c$ drop in the $H$-band, 
compared with those in the $K$-band.
Our models cannot explain such sudden variations, 
as the scattering properties of dielectric particles are less sensitive to
wavelength. 
Since metallic particles show more dependence on the size parameter $x_{eq}$
(Figs.~\ref{anml1}h and i), 
such particles may explain the observation if they exist and are aligned.
An alternative possibility would be contamination by natural light 
in the $H$-band. 
\citet{min91} pointed out that scattered light from Trapezium stars 
and free-free radiation dominate over the scattered light from IRc2 
in the $J$-band. It is possible that a similar situation may occur also 
in the $H$-band to some extent.

\subsection{Degree of Alignment}  \label{degree}
We explore the range of the Rayleigh reduction factor $R$ which can explain 
the linear and circular polarization observed in the $K$ and $L$-bands  
for the CD-model. 
Using the results of calculations in Section~\ref{cal}, 
we examine 4913 ($=17\times17\times17$) models in the 3-D parameter space 
of $\theta_i=0-180\arcdeg$, $\theta_s=0-180\arcdeg$, and 
$\phi_s=0-180\arcdeg$. 
Assuming a value for the Rayleigh reduction factor $R$ 
(column 1 in Table~\ref{tbl-2} and \ref{tbl-3}), 
we count the numbers out of 4913 models which satisfy 
the following conditions:
\begin{equation}
p_c^{obs} - \delta p_c < p_c < p_c^{obs} + \delta p_c, 
\end{equation}
\begin{equation}
p_l^{obs} - \delta p_l < p_l < p_l^{obs} + \delta p_l, 
\end{equation}
and 
\begin{equation}
|PA^{obs} - PA| < \delta PA,
\end{equation}
and the results are shown in column 3 in those tables.
In the $K$-band (Table~\ref{tbl-2}), the observed values 
(and adopted ranges of acceptance within the parentheses) are 
$p_c^{obs}=15(3)\%$, $p_l^{obs}=38(7)\%$, and $PA^{obs}=0(5)\arcdeg$, 
and in the $L$-band (Table~\ref{tbl-3}), those values are 
$p_c^{obs}=9.4(3)\%$, $p_l^{obs}=57(11)\%$, and $PA^{obs}=0(5)\arcdeg$. 
Those adopted ranges of acceptance are larger than 
the observational errors, 
because the latter are too small to yield meaningful results 
with our models. 
The column 2 in Table~\ref{tbl-2} and \ref{tbl-3} is 
the maximum $|p_c|$, and  
the columns 4, 5, and 6 are the averages (and standard deviations
within the parentheses) 
of the angles $\theta_i$, $\theta_s$, and $\phi_s$, respectively,  
with which the model satisfies the above conditions.    

Tables~\ref{tbl-2} and \ref{tbl-3} show that $R$ should be 
larger than $\approx$0.4 or 0.6 to explain the observations. 
These values are much larger than the value of $R \approx 0.25$ 
which is derived for oblate grains with the axial ratio 1.5 by \cite{hilde95}, 
from dichroic linear polarization at $\lambda=2.2\mu$m,  
thermal emission at $100\mu$m, and also from 
the observed 9.7 $\mu$m silicate absorption band feature.
It is also noted that the fraction of models which satisfy 
the observations is small, i.e., $\approx10$ to $80$ out of 4913,
or $\approx10^{-3}$ to $10^{-2}$. 
This fraction will increase if we adopt larger values 
for the ranges of acceptance, though it is still $\approx10^{-1}$
if the ranges are two times larger than the present assumption.
This means that the optimal directions of incidence and scattering
of light are required to explain the observations, and 
that large circular polarization would not always be expected  
even if the alignment is strong.  
Alternatively, it would be necessary to include other light scattering 
processes that are not considered here, i.e., multiple scattering 
and/or dichroic extinction, to fully explain the observations. 
The effects of those processes are discussed in Sections~\ref{pol-inc} 
and \ref{dich-pol}.

Fig.~\ref{shape}a shows the Rayleigh reduction factor $R$ 
necessary to obtain $p_c=15$\% in the $K$-band as a function 
of the axial ratio $r$, 
where the axial ratios for ellipsoids are set as $r:\sqrt{r}:1$. 
We show exact results computed with the CD-model 
(lines and open symbols).  Also shown are results 
from the approximate expression $R=0.15/p_c^{max}$ (crosses and pluses), 
which is based on the fact that circular polarization is approximately 
proportional to $R$ if $mx_{eq} \la 5$ (Section~\ref{precessing}).
The Rayleigh reduction factor $R$ shows a rapid decrease for $r \approx 1$, 
reaches a minimum around $r \approx 2$, and then 
slowly increases with $r$ for $r \ga 2$.
This figure shows again that strong alignment, i.e., $R \ga 0.5$, is 
required to explain the observation, even if the shape is highly elongated 
or flattened. 

How can we explain such large values of $R$? 
We give a few comments here.
The Davis-Greenstein mechanism, which is based on grain rotation 
driven by gas-grain collisions and on paramagnetic relaxation, 
gives a moderately large value of the Rayleigh reduction factor $R$ 
when the ratio of grain temperature to gas temperature 
is significantly small, i.e., $<<10^{-1}$, \citep{roberge99}. 
However, in dense environments such as the OMC1 region, 
the gas and grain temperatures should be nearly equal. 
Thus we cannot expect $R$ to be large with the Davis-Greenstein mechanism. 
The alignment by the difference of velocities between gas and grain, 
originally known as the Gold-mechanism, is expected 
in the presence of Alfv\'enic or magnetosonic waves \citep{laz94,laz97}.
The maximum $R$ expected with this mechanism is $\approx0.25$ 
and is smaller than what our analysis yields. 
The alignment by radiative torques is expected when irregularly,
or helically, shaped grains scatter light \citep{DM76,DW96,LH07,Hoang08}.
For a review of recent alignment theory, see \cite{Laz07}.
Since SEBN is a reflection nebula, the grains in this region 
are strongly illuminated by the star IRc2.
Thus we may expect this mechanism to be working there. 
It is unfortunate that the study of this mechanism is 
beyond the scope of the present paper because we cannot evaluate 
the effects produced by helical grains 
with the FIM nor the Tmat method.

\subsection{Direction of the Alignment} \label{direction}
We have discussed the angles of incidence ($\theta_i'$) 
and scattering ($\theta_s'$ and $\phi_s'$) 
with respect to the reference frame of the grain. 
However, in the astronomical context, it is much more convenient 
to use other angles, i.e., the position angle $PA'$ that is an azimuthal angle 
of the alignment projected on sky, measured from north to east, 
and the scattering angle $\Theta_{sca}$. 
We need one more angle to specify the direction of alignment in space,  
i.e. the inclination angle between the alignment and the line of sight, 
and this angle is the same as $\theta_s$ already defined above. 
We assume that the SEBN region is illuminated by IRc2  
from a direction with position angle $84\arcdeg$,  
i.e. illuminated almost from the west direction. 

With trigonometry, we calculate the angles $PA'$ and $\Theta_{sca}$  
in columns 7 and 8, respectively, in Tables~\ref{tbl-2} (the $K$-band) and 
\ref{tbl-3} (the $L$-band).  
Almost all the models with an axial ratio of 2:1 and the oblate model with 3:1 
show similar results if their size distribution is $a_{eq}=0.15-1.5 \mu$m. 
However, the results of prolate with 3:1 are different from the others, 
suggesting a strong shape dependence for elongated particles. 
The models with smaller size distribution of $a_{eq}=0.1-1 \mu$m 
also show different results, especially in the $L$-band. 
With the exception of these models, the angles of acceptable models are 
$(\theta_i',\theta_s', \phi_s') \approx (60\arcdeg-80\arcdeg,
40\arcdeg-60\arcdeg,120\arcdeg-140\arcdeg)$, 
$PA' \approx 20\arcdeg-50\arcdeg$, and 
$\Theta_{sca} \approx 100\arcdeg-110\arcdeg$. 
The deduced position angle $PA' \approx 20\arcdeg-50\arcdeg$ is 
almost perpendicular to the direction of the magnetic field, 
i.e. $PA'\approx150\arcdeg$, observed in the region of SEBN 
\citep{chrys94}, 
and also to the average direction, $PA'\approx120\arcdeg$, 
in the overall region of OMC1 \citep{chrys94, houde04}. 
This suggests that the magnetic field would vary locally 
in a small region of SEBN where light is scattered, 
if the direction of alignment is parallel to the magnetic field 
as is usually assumed.

\subsection{Effects of Polarized Incident Light} \label{pol-inc}
We now assume that the incident light is polarized, and investigate 
how the polarization status is changed. 
The normalized Stokes parameters $(q'_s, u'_s, v'_s)$, 
where $q'_s=Q'_s/I'_s$ etc., of scattered light for polarized incident 
light are calculated with eq.(\ref{stks}),  
and are compared with those $(q_s, u_s, v_s)$ for nonpolarized 
incident light.
Figs.~\ref{intri}(a) and (b) show the results for $(q_i, u_i, v_i) = (0.3, 0, 0)$, 
Figs.~\ref{intri}(c) and (d) for $(0, 0.3, 0)$, and 
Figs.~\ref{intri}(e) and (f) for $(0, 0, 0.3)$.
We assume oblate grains with an axial ratio of 2:1, 
composed of silicate, and size distribution of $a_{eq}=0.15-1.5\mu$m. 
While the Stokes parameters $q'_s$ and $u'_s$ show significant deviations,
$\approx$30\%, from $q_s$ and $u_s$ (Figs.~\ref{intri}a and c), 
the effect for circular polarization $v'_s$ is less significant, 
$\approx$10\% (Figs.~\ref{intri}b and d).
Since the intensity of scattered light $I'_s$ is approximated by 
$I'_s \approx F_{11}I_i$, the expressions of the scattered light become 
simpler and are given by
\begin{equation}
q'_s \approx f_{21} + f_{22}q_i + f_{23}u_i + f_{24}v_i, 
\end{equation}
\begin{equation}
u'_s \approx f_{31} + f_{32}q_i + f_{33}u_i + f_{34}v_i, 
\end{equation}
and
\begin{equation}
v'_s \approx f_{41} + f_{42}q_i + f_{43}u_i + f_{44}v_i, 
\end{equation}
where $f_{jk} = F_{jk}/F_{11}$. Since $f_{jk}$ varies from -1 to 1, 
we can set upper and lower limits for the deviations, and show them as 
the dashed lines in Fig.~\ref{intri}. 
The quantities $q'_s$ and $u'_s$ go to the limits, 
while the values of $v'_s$ are far smaller. 
Therefore, if the incident circular polarization $v_i \approx0$, 
we may write the deviation of circular polarization 
from the nonpolarized incident model as 
\begin{equation}
|v'_s - v_s| \la 0.3p_{li},
\end{equation}
where $p_{li}$ $(=\sqrt{q_i^2+u_i^2})$ is the degree of 
linear polarization of the incident light.  
The effect of a polarized incident light is 
not significant unless $p_{li}$ is extremely large.

Fig.~\ref{cnt2d} shows contour plots of the number of models
that satisfy the observed linear and/or circular polarization 
in the $K$-band (eqs.(20)-(22) in Section~\ref{degree}) 
out of 4913 models (see Section~\ref{degree}),
for oblate grains with an axial ratio 2:1, composed of silicate,
with size distribution of $a_{eq}=0.15-1.5\mu$m.   
Here we assume that the normalized Stokes parameters of incident light  
$q_i$ and $u_i$ go from $-1$ to 1, and that the circular polarization 
$v_i$ is null. 
The number of models with larger linear (Fig.~\ref{cnt2d}a)
and circular (Fig.~\ref{cnt2d}c) polarization increases  
with larger $q_i$ and $u_i$, 
while the models that explain the position angle $PA$ are restricted 
only to a region of relatively small $q_i$ and $u_i$ (Fig.~\ref{cnt2d}b). 
Therefore, the acceptable models that explain all the observed properties 
are also found in a region of small $q_i$ and $u_i$, 
centered on $(q_i,u_i)=(0.15,-0.27)$ (Fig.~\ref{cnt2d}d).
In this region, the number of acceptable models is about 50, and 
is two times larger than that on $(q_i,u_i)=(0,0)$, 
showing that the presence of optimum polarized incidence is effective 
to explain the observations.
However, it is also noted that very large incident linear polarization, 
i.e. $\ga 50\%$, is not useful. 
Although the effects of polarized incident light are significant, 
large incident linear polarization will strongly affect 
the position angles, not consistent with the observation.

\subsection{Polarization by Dichroic Extinction} \label{dich-pol}
If grains are strongly aligned as discussed in Section~\ref{degree}, 
the transmitted light through space containing those grains would be 
also linearly polarized.
We estimate the ratio of dichroic linear polarization to 
optical depth $p/\tau$ in our models, as a function of the axial 
ratio $r$ of the grains (Fig.~\ref{shape}b).   
We show results obtained with formula $p/\tau = R\cdot (p/\tau)_{per}$, 
where $R$ is the Rayleigh reduction factor necessary to obtain 
$p_c=15$\% in the $K$-band as given by our exact calculations,  
and $(p/\tau)_{per}$ is the value for perfect alignment. 
The value of $(p/\tau)_{per}$ becomes larger with $r$,   
and thus the derived value of $p/\tau$ also becomes larger, 
even if $R$ decreases. 
When $r \approx 2-2.5$, the value of $p/\tau$ is nearly equal to 
the observed maximum $\approx0.07$ (Fig.1 of \citet{jones89}), 
although the exact value of $p/\tau$ in SEBN is not known.

Circular polarization can be produced by extinction 
with aligned nonspherical grains, i.e., dichroic polarization or 
Mechanism 2 explained in Section 1.
The exact solution of the Stokes parameters with this 
mechanism is found in \citet{whit02} and \citet{lucas05}. 
We can write an expression of circular polarization $p_c^d$ as 
\begin{equation}
p_c^d \approx u^i \tau R K_{34}/K_{11}, 
\end{equation}
in the first-order approximation,  
where $u^i$ is a normalized Stokes parameter "$u$" of the incidence light, 
$\tau$ is the optical depth along the line of sight, 
$R$ is the Rayleigh reduction factor, and  
$K_{11}$ and $K_{34}$ are components of the extinction matrix \citep{mht00}.
The value of $K_{34}/K_{11}$ is $\approx 0.1$
in our spheroidal/ellipsoidal silicate grains with radii of 
0.15-1.5 $\mu$m at $\lambda=2.2\mu$m.
If we further assume $u^i \approx 0.3$, $\tau \approx 1$, and 
$R \approx 0.5$, then we obtain $p_c^d \approx 1.5$\%.
Although each parameter is quite uncertain, 
the calculated value in our model is far below the observed 
circular polarization of 15\%.

If the size of grains is much smaller than wavelength, 
i.e. within the Rayleigh approximation, the effect of dichroic polarization
will increase, in contrast to scattering. 
An example is the model of those grains that are assumed in diffuse 
interstellar space, i.e., $a_{max}\approx 0.25\mu$m.
As for linear polarization, 
the value of $(p/\tau)_{per}$ will be 0.28 for oblate silicate grains 
with an axial ratio of 2, and the observed maximum value of 
$p/\tau\approx0.07$ in various clouds \citep{jones89} can be explained 
with $R\approx0.25$.  For circular polarization, 
the value of $K_{34}/K_{11}$ will be much larger than unity \citep{lucas05}.
For dielectric particles, $K_{34}$ is proportional to $a_{eq}$, 
while $K_{11} varies as a_{eq}^{4}$, and thus the ratio $K_{34}/K_{11}$ 
goes as $a_{eq}^{-3}$ in the Rayleigh approximation.
If dielectric small nonspherical grains are aligned, 
then the value of $K_{34}/K_{11}$ would be larger than unity, 
and the observations could be explained with dichroic polarization.
Therefore, on the assumption of smaller grains, one can obtain 
large linear and circular polarization with the mechanism of 
dichroic polarization \citep{lucas05}.

\subsection{The $3\mu$m Ice-band Feature}  \label{band-feature}
Spectropolarimetry of transmitted light of BN object and other IR sources 
shows that the $3\mu$m ice band is linearly polarized \citep{DB74,hough96}.  
This polarization is explained with the accretion of ice on 
aligned spheroidal silicate and/or graphite grains \citep{LD85}.
We assume that such grains with ice are present also in the SEBN region, 
and examine the effects of the $3\mu$m ice band on polarization in 
scattered light. 
In our model, the grains grow in size by accretion of ice, 
and the refractive index is assumed homogeneous and is approximated 
with the Bruggesman mixing rule which is based on an effective 
medium theory \citep{kru03}.
This approximation is adopted because our present codes of FIM or 
Tmat cannot calculate ellipsoidal core-mantle grains. 
We use the refractive index for the "strong ice mixture" with temperature 120K 
from \citet{hudgins93}, which contains $H_2O:CH_3OH:CO:NH_3 = 100:50:1:1$.  
The refractive index for silicate is the same as in the previous sections.

Fig.~\ref{mix-w50} shows the results for oblate grains with an axial ratio 2:1. 
The size of silicate grains without ice is $0.15-1.5\mu$m. 
The degree of linear polarization $p_l$ in the 3$\mu$m ice feature 
increases with the volume of ice, 
while the circular polarization $p_c$ decreases. 
In the dichroic extinction model \citep{aitken06},
both the degrees of $p_l$ and $p_c$ increase in the feature. 
The difference is due not only to the light scattering process itself, 
but also to the difference of grain size in the models, 
i.e. grains are larger in our model. 
Therefore circular polarization in the 3$\mu$m band is 
a useful diagnostic for finding out which process is working 
in the SEBN region.

\section{Conclusions}

We have studied polarization in scattered light by imperfectly aligned
spheroidal or ellipsoidal grains with FIM and Tmat.  
The shapes of grains considered here include tri-axial ellipsoid 
which has not been investigated before.
Our main conclusions are as follows:

1. With using spherical harmonics and associated Legendre polynomials, 
we have investigated the angular dependence of circular polarization 
$p_c$ or the component $F_{41}$ of Mueller scattering matrix. 
For dielectric grains 
that are aligned spinning around shortest axis, 
the angular dependence of $p_c$ or $F_{41}$ does not 
vary much in different shapes, if the grains are not much elongated and 
not large, i.e., the axial ratio of grains is up to about 2:1 
and $|m|x_{eq}$ is up to $\approx 5$.  For those grains, 
$p_c$ is approximately proportional to the Rayleigh reduction factor $R$, 
even when the scattering properties are far from the Rayleigh approximation, 
i.e. $|m|x_{eq}\approx 3-5$.

2. To explain the large linear and circular polarization 
observed in SEBN of OMC1 not only in the $K$, but also in the $L$-band,
the size distribution of silicate grains should range from $0.15$ to 
$1.5\mu$m; these sizes are larger than those assumed in previous papers.  
With those grains, we deduce the Rayleigh reduction factor $R \ga0.5$ 
in the SEBN region.
Such a strong alignment cannot be explained by the Davis-Greenstein mechanism. 
We suggest alignment by radiative torques as an alternative mechanism. 
We also investigate possible orientations of grain alignment, 
and those of incident and scattered beams in our models. 
The orientations should be almost optimal, and this restricts 
possible configurations significantly. 

3. We investigate how the circular polarization in scattered light
is affected by linear polarization $p_{li}$ in incident light. 
The effect of linearly polarized incident light is small, i.e., 
the difference between circular polarization for linearly 
polarized incident light $p_{li}$ and 
that for nonpolarized incidence is less than $0.3p_{li}$ 
in our models. 
This result shows that the conversion from linear to circular 
polarization would not be a dominant process to produce large 
circular polarization, unless $p_{li}$ is extremely large.
Also the effect of dichroic polarization is small in our models, 
although it will be more significant than scattering 
if the grain sizes are less than those assumed here.

4. If the grains are composed of silicates and ice, the effect of 
the $3\mu$m ice band should appear in polarization of scattered light.
In our models, the degree of linear polarization increases  
while that of circular polarization decreases in the $3\mu$m band. 
Linear and circular polarimetry of the $3\mu$m ice band in the SEBN region
should be important to investigate the details of the scattering 
process.

%% In this section, we use  the \subsection command to set off
%% a subsection.  \footnote is used to insert a footnote to the text.


%% This section also includes several of the displayed math environments
%% mentioned in the Author Guide.


%% Included in this acknowledgments section are examples of the
%% AASTeX hypertext markup commands. Use \url without the optional [HREF]
%% argument when you want to print the url directly in the text. Otherwise,
%% use either \url or \anchor, with the HREF as the first argument and the
%% text to be printed in the second.

\acknowledgments

We thank the referee Philip Lucas for constructive comments. 
We also thank Alex Lazarian for comments on grain alignment. 
We are grateful to the Japanese Society for the Promotion of Science (FY2002), 
the Kagawa University International Exchange Foundation (FY2003), 
the Kagawa University Specially Promoted Research Fund (FY2008), 
and the Natural Sciences and Research Council of Canada 
for supporting this research.
We have used the Fortran program of Tmat "ampld.new.f" dated 04/03/2003 
written by Mishchenko (\url{http://www.giss.nasa.gov/}\verb|~|\url{crmim}). 
The refractive index of smoothed astronomical silicate is 
cited in Jena-St.Petersburg Database of Optical Constants (JPDOC, 
\url{http://www.astro.uni-jena.de/Laboratory/Database/jpdoc/index.html}).

% More information on the AASTeX macros package is available \\ at
% \url{http://www.aas.org/publications/aastex}.
% For technical support, please write to
% \email{aastex-help@aas.org}.


%% copy editing.  Individual instruments can be provided in parentheses,
%% after the keyword, but they will not be verified.

% Facilities: \facility{Nickel}, \facility{HST(STIS)}, \facility{CXO(ASIS)}.



\appendix

\section{Sign of the Stokes Parameters}

We use the same definition of polarization as in \citet{vdH57}, 
i.e., the circular polarization is positive ($V>0$) 
when the electric vector is rotating clockwise with time as seen 
from an observer. 
However, the time component of the electric and magnetic fields in 
\citet{vdH57}, i.e., $e^{+i \omega t}$, is different 
from that in the FIM and Tmat calculations ($e^{-i \omega t}$). 
The Stokes parameters are thus written for incident light as
\begin{equation}
    I_i =  E_{i2} E_{i2}^*  + E_{i1} E_{i1}^*, 
\end{equation}
\begin{equation}
    Q_i  =  E_{i2} E_{i2}^*  - E_{i1} E_{i1}^*,
\end{equation}
\begin{equation}
    U_i  =  E_{i2} E_{i1}^*  + E_{i1} E_{i2}^*,
\end{equation}
\begin{equation}
    V_i  =-i(E_{i2} E_{i1}^*  - E_{i1} E_{i2}^*),
\end{equation}
where $E_{i1}$ and $E_{i2}$ are amplitudes of orthogonal electric vectors. 
The sign of $V_i$ in eq.(A4) is different from that by \citet{vdH57} (p.41). 
The Stokes parameters for scattered light are also written in the same manner. 
The expressions for transformation matrix $F_{jk}$ of the Stokes 
parameters (eq.\ref{stks}) are different from those in \citet{vdH57} (p.44) 
for the sign in the fourth line and in the fourth column, except for $F_{44}$.
This definition of the Stokes parameter $V$ is different from that in 
our previous papers \citep{mb04,bm05}, but is same as that of \citet{gled00} 
and other papers.

\section{Expression of $F_{41}$ for a Precessing Grain}

When the grain is small, or in the Rayleigh approximation, 
the component $F_{41}$ of the Mueller matrix is expressed 
only with the first term $f^{112}$ of eq.(\ref{F41}) or eq.(\ref{F41-2}):
\begin{equation}
F_{41}(\theta_i,\theta_s,\phi_s) = a_{112}f^{112}(\theta_i,\theta_s,\phi_s) =
a_{112} \cdot \frac{3}{4}\sqrt{\frac{5}{\pi}}\sin\theta_i\cos\theta_i\sin\theta_s\sin\phi_s.
\end{equation}
We now derive the expression of $f^{112}$ when the grain is precessing (S-model) 
around the direction of $B$, and prove eq.(\ref{R}).
We first rewrite eq.(B1) with the dashed angles of Fig.~\ref{sph2}, 
which are based on the direction of alignment $B$, 
and then integrate it over $\phi_a'$ from 0 to $2\pi$. 
The term $f^{112}$ can be written as 
\begin{equation}
f^{112} \propto
\sin\theta_i\cos\theta_i\sin\Theta_{sca}(\sin\angle BIS \cos\angle BIA - \cos\angle BIS \sin\angle BIA),
\end{equation}
where we have used the relation
\begin{equation}
\sin\theta_s\sin\phi_s=\sin\Theta_{sca}\sin(\angle BIS - \angle BIA),
\end{equation}
for the spherical triangle $IAS$. We further use the following relations: 
\begin{equation}
\sin\Theta_{sca} \sin\angle BIS = \sin\theta_s' \sin\phi_s',
\end{equation}
for the spherical triangle $BSI$, and
\begin{equation}
\sin\theta_i \sin\angle BIA = \sin\theta_a' \sin\phi_a',
\end{equation}
\begin{equation}
\sin\theta_i \cos\angle BIA = \cos\theta_a' \sin\theta_i' - \sin\theta_a' \cos\theta_i' \cos\phi_a',
\end{equation}
\begin{equation}
\cos\theta_i  = \cos\theta_a' \cos\theta_i' + \sin\theta_a' \sin\theta_i' \cos\phi_a',
\end{equation}
for the spherical triangle $BIA$.

With eqs.(B4)-(B7), eq.(B2) can be  rewritten as
\begin{equation}
f^{112} \propto (A-B\cos\phi_a'-C\sin\phi_a')(D+E\cos\phi_a')
\end{equation}
where
\begin{equation}
A = \cos\theta_a' \sin\theta_i' \sin\theta_s' \sin\phi_s',
\end{equation}
\begin{equation}
B = \sin\theta_a' \cos\theta_i' \sin\theta_s' \sin\phi_s',
\end{equation}
\begin{equation}
C = \sin\theta_a' \sin\Theta_{sca} \cos\angle BIS,  
\end{equation}
\begin{equation}
D = \cos\theta_i' \cos\theta_a',
\end{equation}
\begin{equation}
E = \sin\theta_i' \sin\theta_a', 
\end{equation}
which are constant at present.
If we integrate $f^{112}$ over $\phi_a'$ from 0 to $2\pi$, 
only the constant term and the term with $\cos^2\phi_a'$ 
in eq.(B8) remain, and other terms vanish.
We thus finally obtain the average $f^{112}$ as 
\begin{equation}
<f^{112}>  \propto  \sin\theta_i' \cos\theta_i' \sin\theta_s' \sin\phi_s'(1-(3/2)\sin^2\theta_a').
\end{equation}
Comparing eq.(B14) with eq.(B1), we see the angular dependence is 
the same for $\theta_i'$, $\theta_s'$, and $\phi_s'$,
but the amplitude is different by a factor of $(1-(3/2)\sin^2\theta_a')$, 
which is the same as the Rayleigh reduction factor $R$ (see eq.(\ref{defR})).  
Therefore, we have shown that $<f^{112}>$ is proportional to $R$, 
i.e. eq.(\ref{R}) in Section~\ref{precessing}. 

The second term $f^{211}$ in eq.(\ref{F41}) or eq.(\ref{F41-2}) is 
also proportional to the Rayleigh reduction factor. 
One can prove it in the same manner as for $f^{112}$.




\clearpage

%% the file, uncomment the \plotone, \plottwo, and
%% If you need a layout that cannot be achieved with \plotone or
%% \plottwo, you can invoke the graphicx package directly with the
%% \includegraphics command or use \plotfiddle. For more information,
%% please see the tutorial on "Using Electronic Art with AASTeX" in the
%% documentation section at the AASTeX Web site,
%% http://www.journals.uchicago.edu/AAS/AASTeX.
%%
%% The examples below also include sample markup for submission of
%% supplemental electronic materials. As always, be sure to check
%% the instructions to authors for the journal you are submitting to
%%for specific submissions guidelines as they vary from

%% This example uses \plotone to include an EPS file scaled to
%% 80% of its natural size with \epsscale. Its caption
%% has been written to indicate that additional figure parts will be
%% available in the electronic journal.

\begin{figure}
\epsscale{1.0}
\plotone{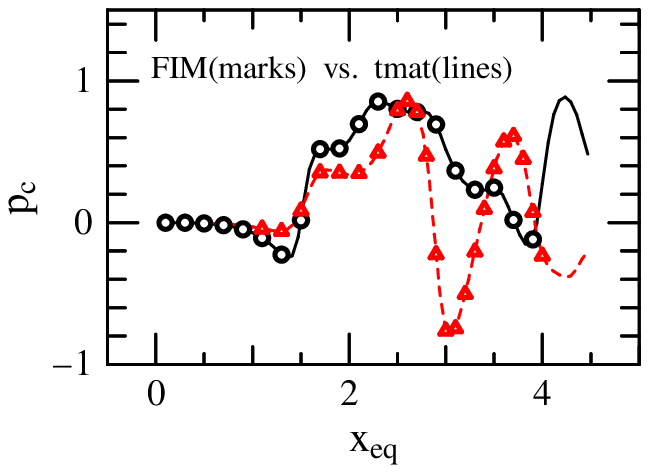}  %% \plotone{f_vs_t.eps}   %%% Fig.1 %%
\caption{Comparison of the results computed with FIM (symbols) and 
with Tmat(lines) for an oblate particle (axial ratio is 2:1 and $m=1.7$). 
The circles (FIM) and the solid line (Tmat) are results 
for $(\theta_i,\theta_s,\phi_s)=(60\arcdeg, 60\arcdeg, 135\arcdeg)$, and 
the triangles (FIM) and the broken line (Tmat) are those 
for $(\theta_i,\theta_s,\phi_s)=(60\arcdeg, 29\arcdeg, 135\arcdeg)$.  
\label{f_vs_t}}
\end{figure}

\begin{figure}
\epsscale{1.0}
\plotone{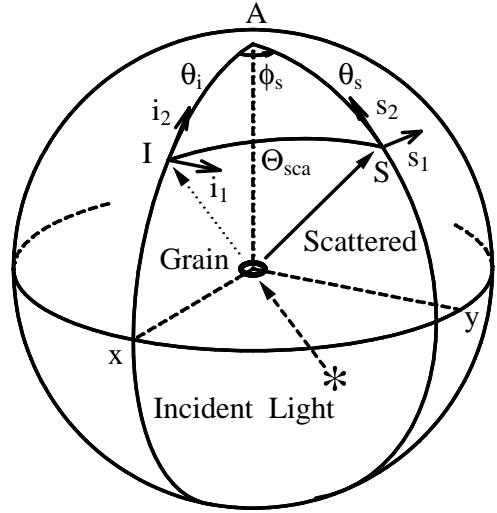}  % \plotone{SPH1.EPS}     %%% Fig.2 %%
\caption{Configuration of grain relative to the incident and scattered 
light beams.
The grain is at the origin and the light source is within the plane of 
$y=0$ (and $z<0$ in this figure).  The incident light goes 
in the direction of $I$ within the xz-plane and its direction is defined 
by the angle $\theta_i$ 
while the scattered light going in the direction of $S$ is defined 
by the angles $\theta_s$ and $\phi_s$. 
The scattering angle $\Theta_{sca}$ is the angle between $I$ and $S$.
\label{sph1}}
\end{figure}

\begin{figure}
\epsscale{1.0}
\plotone{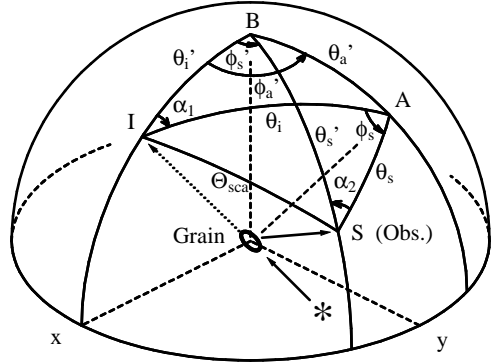}   % \plotone{SPH2.EPS}     %%% Fig.3 %%
\caption{Geometry of the problem showing the grain configuration 
with respect to the directions of the incident $I$ and 
scattered $S$ light beams for a precessing grain. 
$A$ is a direction related to the symmetry of the grain and 
$B$ the direction of alignment. See the text for more details.
\label{sph2}}
\end{figure}

\begin{figure}
\epsscale{1.00}
\plotone{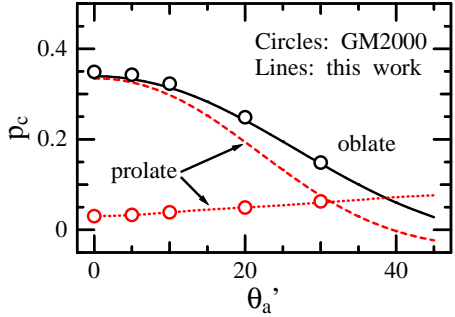}   % \plotone{gm.eps}       %%% Fig.4 %%
\caption{Comparison the circular polarization $p_c$ from \cite{gled00} 
(circles) and that computed here (lines) for spheroidal silicate grains 
(m=1.71+0.03i, axial ratio 2:1). 
The calculations were performed at a wavelength of 1 $\mu$m 
for a size distribution defined by minimum and maximum grain sizes of 
0.1-1.0 $\mu$m and a power law with an index of $-3.5$. 
For oblate models, the incident and scattered directions are 
set to $(\theta_i',\theta_s',\phi_s')=(72.0\arcdeg, 72.9\arcdeg, 95.7\arcdeg)$.
For prolate models they are 
$(\theta_i',\theta_s',\phi_s')=(50.0\arcdeg, 45.6\arcdeg, 145.3\arcdeg)$ 
for open circles and dotted line,  and 
$(\theta_i',\theta_s',\phi_s')=(75.5\arcdeg, 82.8\arcdeg, 112.5\arcdeg)$ 
for dashed line.
\label{gm}}
\end{figure}

\begin{figure}
\epsscale{1.00}
\plotone{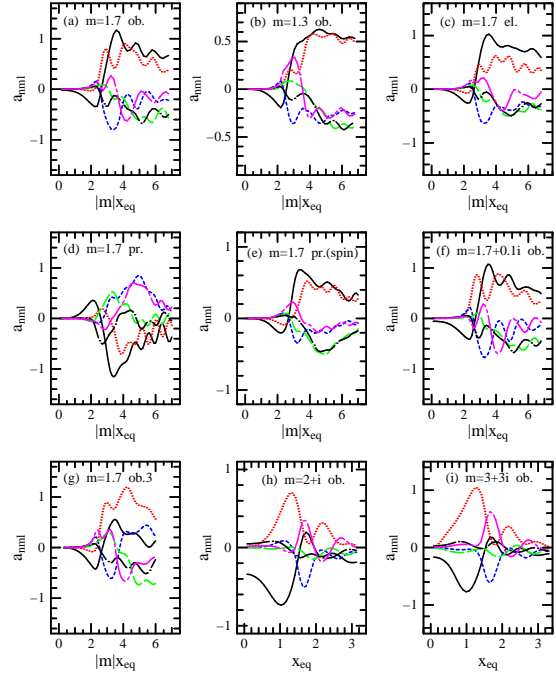}   % \plotone{M17-11.EPS}   %%% Fig.5 %%  ~/figs/ice_sil/M17-11.EPS %%%
\caption{Dependence of $a_{nml}$ on $|m|x_{eq}$ or $x_{eq}$ for 
(a) an oblate grain with an axial ratio 2:1 and $m=1.7$,  
(b) same as (a) but for $m=1.3$,
(c) same as (a) but for an ellipsoidal grain with axial ratios $2:\sqrt{2}:1$,
(d) same as (a) but for a prolate grain, 
(e) same as (a) but for a spinning prolate grain, 
(f) same as (a) but for $m=1.7+0.1i$,
(g) same as (a) but for an oblate grain with axial ratio 3:1,
(h) same as (a) but for $m=2+1i$, and 
(i) same as (a) but for $m=3+3i$.  
Solid line is for $(n,m,l)= (1,1,2)$, 
dotted line $(2,1,1)$, 
broken line $(2,2,2)$,
long dashed line $(2,1,3)$,
dotted chain $(3,1,2)$, and 
dashed chain $(3,2,1)$.
\label{anml1}}
\end{figure}

\begin{figure}                 %% Fig.6 %%
\epsscale{1.00}
\plotone{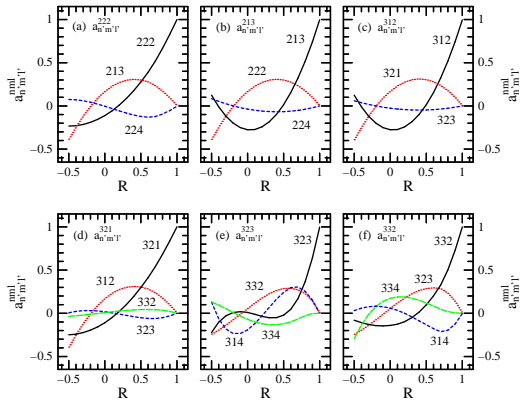}   % \plotone{anml.eps}  %% from fim21/sph_har/222/anml.eps % NOT in ~/figs/anml %%
\caption{Dependence of $a_{n'm'l'}^{nml}$ on $R$. 
The values of $(n,m,l)$ are 
(a) $(2,2,2)$,  (b) $(2,1,3)$,  (c) $(3,1,2)$,  (d) $(3,2,1)$,  
(e) $(3,2,3)$, and (f) $(3,3,2)$.
In each graph, various curves are given for different values of 
$n',m',l'$ as indicated. \label{anml2}}
\end{figure}

\begin{figure}                 %% Fig.7 %%
\epsscale{1.00}
\plotone{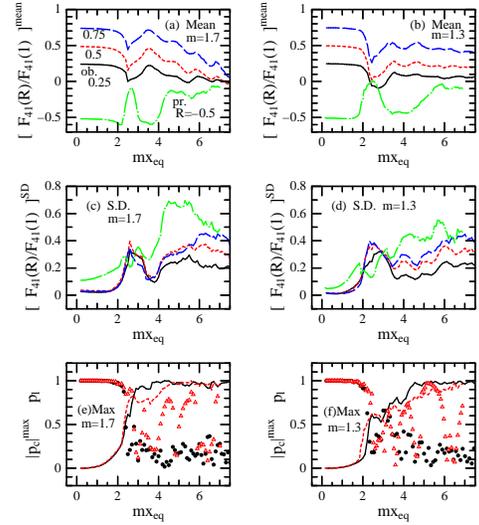}   % \plotone{cmp17-13.eps}         %% comp/cmp17-13.eps %%
% \plottwo{comp.eps}{comp13.eps}
\caption{The effects of grain precession on circular and linear 
polarization. 
(a) The means of $F_{41}(R) / F_{41}(1)$ for $m=1.7$ 
for 1785 or less directions of incidence and scattering (see the text).
Values of the Rayleigh reduction factor $R$ are 0.25 (solid), 0.5 (broken) 
and 0.75 (long dashed) for oblate grains (axial ratio 2:1), 
and -0.5 (dotted chain) for prolate grains (axial ratio 2:1)(see the text).
(b) Same as (a) but for $m=1.3$.
(c) Same as (a) but for the standard deviations.
(d) Same as (c) but for $m=1.3$.
(e) The maximum of absolute values of the circular polarization 
$p_c$ for oblate (solid line) and for prolate (broken line) grains, 
and the linear polarization $p_l$ at the maximum $|p_c|$ 
for oblate (circles) and prolate (triangles) grains with $m=1.7$ and $R=1$.
(f) Same as (e) but for $m=1.3$.  \label{cmp17-13}}
\end{figure}

\begin{figure}                 %% Fig.8 %%
\epsscale{1.00}
\plotone{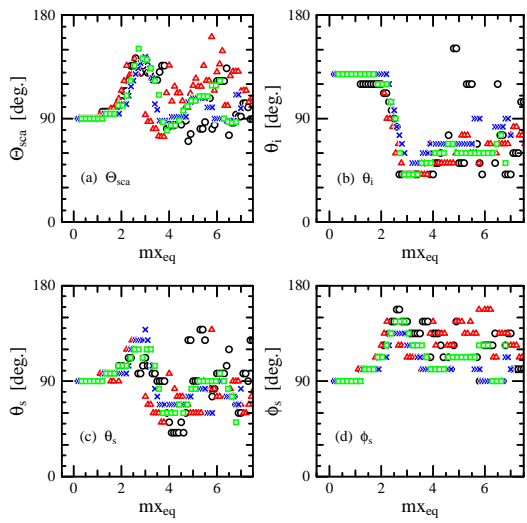}   % \plotone{ANGLES2.EPS}           %% angles2/angles2.eps %%
%%% \plotone{f8.eps}   % \plotone{ANGLES.EPS}           %% angles/angles.eps %%
\caption{The angles for which $|p_c|$ takes its maximum value. 
(a) The scattering angle $\Theta_{sca}$ which corresponds to maximum values 
of $p_c$ for oblate grains with $m=1.7$ (circles), 
oblate grains with $m=1.3$ (triangles), 
spinning prolate grains with $m=1.7$ (crosses), and 
spinning ellipsoidal grains with $m=1.7$ (squares).
(b) Same as (a) but for $\theta_i$.
(c) Same as (a) but for $\theta_s$.
(d) Same as (a) but for $\phi_s$.
The axial ratios are 2:1 for oblate and prolate grains, 
and $2:\sqrt{2}:1$ for ellipsoids.
\label{angles}}
\end{figure}

\begin{figure}       %% Fig.9 %%
\epsscale{1.00}
\plotone{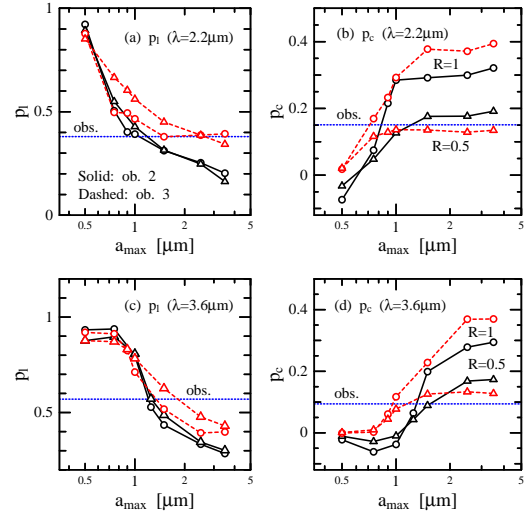}   % \plotone{size.eps}   %% size/size.eps %%
\caption{Linear polarization $p_l$ and circular polarization $p_c$ 
by oblate vs. the maximum size $a_{max}$.  
(a) $p_l$ at at $\lambda = 2.2\mu$m for axial ratio 2:1 (solid line),
and 3:1 (dashed line),  with $R=1$ (circles) and $R=0.5$ (triangles).
(b) Same as (a) but for $p_c$.  
(c) Same as (a) but for $\lambda = 3.6\mu$m.
(d) Same as (c) but for $p_c$.
The values of polarization for axial ratio 2:1 are averages  
within $(\theta_i,\theta_s,\phi_s) = 
(60\pm10\arcdeg, 60\pm10\arcdeg, 135\pm10\arcdeg)$. 
Those for axial ratio 3:1 are for $(\theta_i,\theta_s,\phi_s) = 
(90\pm10\arcdeg, 50\pm10\arcdeg, 100\pm10\arcdeg)$.
These directions are near to those in which $p_c$ shows its maximum 
for $\lambda=2.2\mu$m.
\label{size}}
\end{figure}

\begin{figure}
\epsscale{1.00}
\plotone{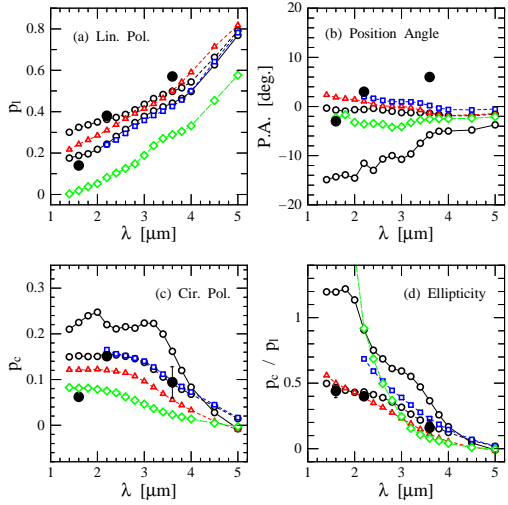}   % \plotone{wv-dep2.eps}      %% Fig.10 %%
\caption{Wavelength Dependence of silicate grains with size distribution 
of $a_{eq}=0.15-1.5\mu$m for (a) degree of linear polarization $p_l$, 
(b) position angle $PA$, (c) degree of circular polarization $p_c$, and 
(d) ellipticity $p_c/p_l$. 
The solid and dashed lines with open circles are for $R=1.0$ and 0.5, respectively,
for oblate grains with an axial ratio 2:1 and $(\theta_i,\theta_s,\phi_s) =
 (60\pm10\arcdeg,50\pm10\arcdeg,140\pm10\arcdeg)$. 
% The broken line without symbol is same but for 
% $(\theta_i,\theta_s,\phi_s)= (90\pm10\arcdeg,150\pm10\arcdeg,140\pm10\arcdeg)$  
% and $R=1$ where the sign of $p_c$ is reversed. 
The broken line with open triangles is for prolate grains 
with an axial ratio 2:1 for $R=0.5$ and $(\theta_i,\theta_s,\phi_s)=
(50\pm10\arcdeg,90\pm10\arcdeg,110\pm10\arcdeg)$.
The dashed line with open squares is for ellipsoidal grains 
with an axial ratio $2:\sqrt{2}:1$ 
for $R=0.5$ and $(\theta_i,\theta_s,\phi_s)=
(70\pm10\arcdeg,50\pm10\arcdeg,120\pm10\arcdeg)$.
The broken line with open diamond is for oblate grains 
with an axial ratio 1.5:1 for $R=0.25$ and 
$(\theta_i,\theta_s,\phi_s)= (70\pm10\arcdeg,70\pm10\arcdeg,140\pm10\arcdeg)$.
The observed values are plotted with filled circles.  \label{wv-dep2}}
\end{figure}

\begin{figure}
\epsscale{1.0}
\plotone{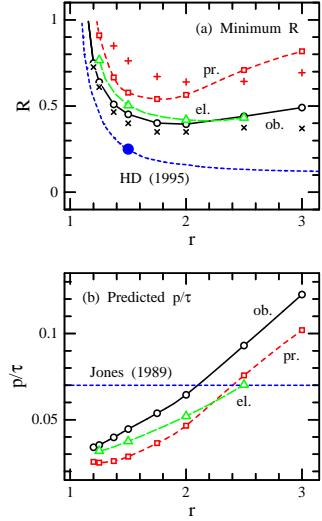}   % \plotone{shape.eps}     %% Fig.11 %%
\caption{The effects of grain shape on polarization at $\lambda = 2.2 \mu$m.
(a) The Rayleigh reduction factor $R$ necessary to obtain $p_c=15$\%. 
The open symbols are the results of exact calculations, 
while crosses and pluses are those for oblate and prolate grains respectively, 
computed with an approximate formula (see the text).
The short dashed line is derived from dichroic extinction and thermal emission 
by \citet{hilde95}, and the filled circle is their preferred value $R=0.25$.
(b) The polarization efficiency $p/\tau$ when the maximum circular 
polarization $p_c$ of 15\% is obtained.  
For ellipsoidal grains, the axial ratios are set as $1:\sqrt{r}:r$ 
where $r$ is the ratio of the maximum and the minimum radii. 
The dotted line is the observed maximum ratio $p/\tau$ \citep{jones89}.
See text for more details. 
\label{shape}}
\end{figure}
% \clearpage

\begin{figure}
\epsscale{1.0}
\plotone{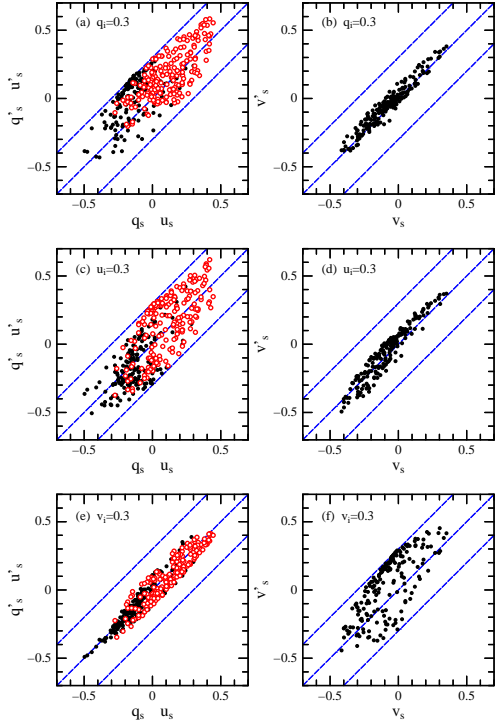}      %% Fig.12 %%   % ~/mb1_more/q1_vs_q2_reduce/p0_30.eps
\caption{The effects of polarization ($q_i$, $u_i$, and $v_i$) 
in incident light on that of the scattered light 
for silicate grains: oblate with an axial ratio of 2:1, 
size distribution of $a_{eq}=0.15-1.5\mu$m, 
the Rayleigh reduction factor $R=0.5$, and wavelength $\lambda = 2.2 \mu$m.
(a) Stokes parameters for linear polarization ($q'_s$ and $u'_s$) of 
scattered light for polarized incidence with ($q_i,u_i,v_i)=(0.3, 0, 0)$
are compared with those ($q_s$ and $u_s$) 
for nonpolarized incident light ($q_i=u_i=v_i=0$).
The filled circles show the stokes parameter of $q_s$ and $q'_s$, 
and open ones $u_s$ and $u'_s$.
(b) Same as (a) but for circular polarization $v_s$ and $v'_s$. 
(c) Same as (a) but for ($q_s,u_s,v_s$)=(0, 0.3, 0).
(d) Same as (c) but for circular polarization $v_s$ and $v'_s$.
(e) Same as (a) but for ($q_s,u_s,v_s$)=(0, 0, 0.3).
(f) Same as (e) but for circular polarization $v_s$ and $v'_s$.
\label{intri}}
\end{figure}
% \clearpage

\begin{figure}
% \epsscale{1.0}
\plotone{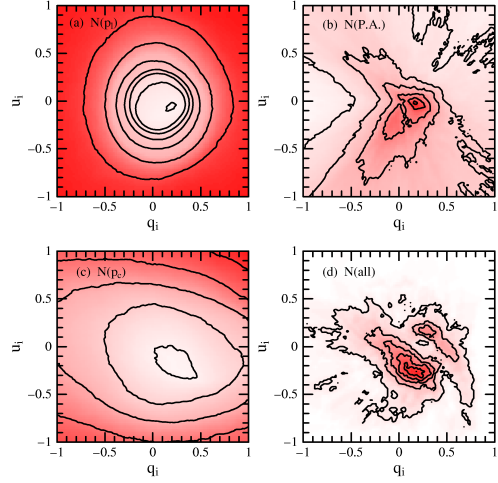}   %% Fig.13 %%  % crab:~/mkcnt2d/N.EPS-pg1.eps (090223)
\caption{Contour plots of the number of models 
that satisfy the observed conditions, 
as a function of linear polarization ($q_i$, $u_i$) 
in the incident light. 
Silicate grains are assumed: oblate with an axial ratio 2:1, 
size distribution of $a_{eq}=0.15-1.5\mu$m, 
the Rayleigh reduction factor $R=0.7$, and wavelength $\lambda = 2.2 \mu$m.
(a) Contour plots out of 4913 models for which $p_l > 31$\%.  
Contour levels are 800, 1200, 1600, ... and  3200. 
The darker areas show larger numbers.
(b) Same as (a) but for $-5\arcdeg < PA <5$\arcdeg. Contour levels are 
400, 500, ... 900.
(c) Same as (a) but for $p_c > 15$\%. Contour levels are
800, 1200, ... 2400. 
(d) Same as (a) but for 
$31\% < p_l < 45\%$, $-5\arcdeg < PA <5\arcdeg$, and $12\% < p_c < 18\%$. 
Contour levels are 5, 15, ..., 55.
\label{cnt2d}}
\end{figure}
% \clearpage

\begin{figure}
\epsscale{1.0}
\plotone{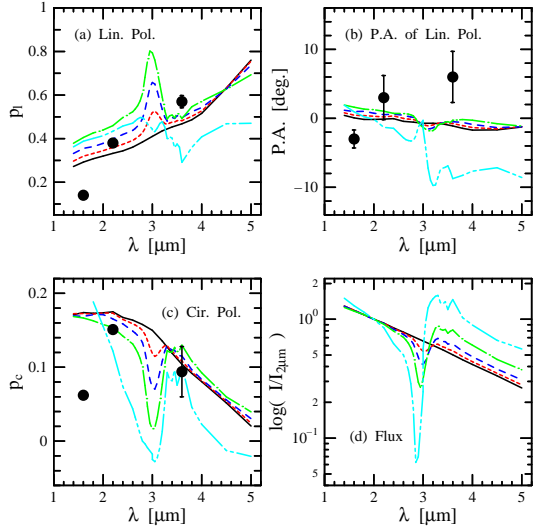}   % \plotone{mix-w50.eps}      %% Fig.14 %%
\caption{The $3\mu$m ice band feature for oblate grains 
with an axial ratio of 2:1 for (a) degree of linear polarization $p_l$, 
(b) position angle $PA$, (c) degree of circular polarization $p_c$, and 
(d) flux normalized at $2\mu$m. 
The solid line is for grains composed of bare silicate
with a size distribution of $0.15-1.5\mu$m, $R=0.5$, and 
$(\theta_i,\theta_s,\phi_s)= (60\pm10\arcdeg,50\pm10\arcdeg,130\pm10\arcdeg)$. 
The short dashed line is the same but for a silicate-ice mixture 
whose volume of ice is 10\%.
The long dashed line is the same but for an ice volume of 25\%.
The dotted chain is the same but for an ice volume of 50\%.
The double dotted chain is for pure ice grains with size distribution 
of $0.193-1.93\mu$m, $R=0.7$, and $(\theta_i,\theta_s,\phi_s)=
(75\pm10\arcdeg,70\pm10\arcdeg,140\pm10\arcdeg)$.
The observed values are plotted with filled circles.  \label{mix-w50}}
\end{figure}
\clearpage

\begin{deluxetable}{ccc}
% \tabletypesize{\scriptsize}
% \rotate
\tablecaption{Conditions for $F_{41}$ or $p_c=0$\label{tbl-1}}
\tablewidth{0pt}
\tablehead{ \colhead{$\theta_i$} & \colhead{$\theta_s$} & \colhead{$\phi_s$} }
\startdata
$0\arcdeg$ or $180\arcdeg$ & any                        & any   \\
any                        & and                        & $0\arcdeg$ or $180\arcdeg$ \\
any                        & $0\arcdeg$ or $180\arcdeg$ & any   \\
$90\arcdeg$                & $90\arcdeg$ or any$^a$     & any   \\
\enddata
\tablenotetext{a}{For the Rayleigh scattering, $\theta_s$ can be any.}
\end{deluxetable}
\clearpage

\begin{deluxetable}{ccrrrrrr}
\tabletypesize{\scriptsize}
% \rotate
\tablecaption{Models that Explain Observations in the K-band$^a$ \label{tbl-2}}
\tablewidth{0pt}
\tablehead{
\colhead{$R$} & \colhead{$|p_c|^{max}$} & \colhead{$N$} & 
\colhead{$\theta_i'$} & \colhead{$\theta_s'$} & \colhead{$\phi_s'$} & 
\colhead{$PA'$} & \colhead{$\Theta_{sca}$}
}
\startdata
%% from crab:/home/matsu/tmat01/silLRG2/pwr35/redf2/15 
\sidehead{Oblate 2:1, L}
 0.2 & 0.07 &   0 & -      & -      & -      & -      & -      \\
 0.4 & 0.15 &  14 & 60( 8) & 56( 6) &136( 8) & 46( 7) &104( 9) \\
% 0.5 & 0.20 &  25 & 62(11) & 50( 8) &137(10) & 46( 9) &101(12) \\
 0.6 & 0.24 &  23 & 63(12) & 48(11) &140(12) & 49(10) &102(15) \\
% 0.7 & 0.29 &  25 & 61(15) & 47(15) &141(13) & 50(12) &100(18) \\
 0.8 & 0.33 &  21 & 67(15) & 41(14) &144(13) & 50(12) &101(19) \\
% 0.9 & 0.38 &  14 & 64(17) & 46(19) &150( 8) & 56( 9) &105(23) \\
 1.0 & 0.43 &  11 & 70(29) & 51(27) &152(12) & 55(16) &115(36) \\
%% from crab:/home/matsu/tmat01/silLRG2o3/pwr35/redf2/15
\sidehead{Oblate 3:1, L}
 0.4 & 0.12 &   0 & -      & -      & -      & -      & -      \\
% 0.5 & 0.15 &  33 & 84(14) & 54( 7) &124(10) & 20(12) &113(12) \\
 0.6 & 0.19 &  31 & 80(16) & 52( 7) &128(11) & 27(13) &112(14) \\
% 0.7 & 0.23 &  12 & 79(17) & 46( 5) &134(15) & 35(15) &112(17) \\
 0.8 & 0.28 &   9 & 83(22) & 43( 9) &135(13) & 35(15) &113(21) \\
% 0.9 & 0.33 &   9 & 97(25) & 42(10) &136(20) & 27(24) &125(24) \\
 1.0 & 0.41 &   7 &125(20) & 32( 8) &114(27) & -7(26) &132(21) \\
%% from crab:/home/matsu/tmat01/silLRG/pwr35/redf 
\sidehead{Oblate 2:1, S}
 0.4 & 0.12 &   0 & -      & -      & -      & -      & -      \\
 0.6 & 0.19 &  31 & 72(12) & 49( 7) &124(10) & 31(10) &102(12) \\
 0.8 & 0.26 &  12 & 68(17) & 41( 7) &136(12) & 43(11) & 99(17) \\
 1.0 & 0.33 &  10 & 67(15) & 38( 6) &138(13) & 46(12) & 97(15) \\
%%% \cline{1-8}
%% from crab:/home/matsu/tmat01/silLRG2pr/pwr35/redf2/15
\sidehead{Prolate 2:1, L}
% \sidehead{Prolate 2, $a_{eq}=0.15-1.5\mu$m, $\lambda$=2.2$\mu$m}
 0.4 & 0.11 &   0 & -      & -      & -      & -      & -      \\
% 0.5 & 0.13 &   3 & 60( 8) & 60( 8) &135( 0) & 45( 5) &106( 9) \\
 0.6 & 0.16 &   8 & 63(10) & 56( 9) &128( 6) & 38( 7) &102(11) \\
% 0.7 & 0.19 &  13 & 58(15) & 62(15) &124( 8) & 38(11) & 99(15) \\
 0.8 & 0.23 &  14 & 59(16) & 61(16) &121( 8) & 36(11) & 98(16) \\
% 0.9 & 0.28 &  18 & 61(18) & 62(18) &116( 9) & 32(14) & 96(17) \\
 1.0 & 0.35 &  20 & 67(26) & 60(26) &110(11) & 24(19) & 94(21) \\
%% /home/matsu/tmat01/silLRG2pr3/pwr35/redf2/15
\sidehead{Prolate 3:1, L}
 0.6 & 0.08 &   0 & -      & -      & -      & -      & -      \\
 0.8 & 0.14 &  10 & 72( 8) & 72( 8) & 90( 0) & 12( 8) & 84( 3) \\
% 0.9 & 0.21 &  32 & 79(12) & 70(13) & 87( 7) &  4(11) & 84( 8) \\
 1.0 & 0.33 &  18 & 78(12) & 75(15) & 94(13) &  7(11) & 91(13) \\
%% from crab:/home/matsu/tmat01/silLRGpr/pwr35/redf
\sidehead{Prolate 2:1, S}
 0.6 & 0.11 &   0 & -      & -      & -      & -      & -      \\
 0.8 & 0.15 &  47 & 69(11) & 66(10) &107( 9) & 20(10) & 96(10) \\
 1.0 & 0.21 &  83 & 75(16) & 62(16) &108(12) & 16(14) & 98(14) \\
%%% \cline{1-8}
%  \\ \\ \\ \\
%% from crab:/home/matsu/fim21/silLRG2el/pwr35/redf2
\sidehead{Ellipsoid $2:\sqrt{2}:1$, L}
 0.4 & 0.14 &   0 & -      & -      & -      & -      & -      \\
% 0.5 & 0.18 &   1 & 41( 0) & 68( 0) &135( 0) & 56( 0) & 99( 0) \\
 0.6 & 0.22 &   5 & 49( 8) & 56(10) &140( 6) & 54( 5) & 96(11) \\
% 0.7 & 0.26 &   7 & 53(12) & 52(13) &138( 9) & 52( 8) & 95(15) \\
 0.8 & 0.31 &   8 & 69(14) & 38(10) &132(12) & 40(11) & 96(15) \\
% 0.9 & 0.36 &  12 & 62(22) & 47(26) &131(11) & 42(12) & 97(27) \\
 1.0 & 0.41 &  21 & 59(27) & 57(32) &131(13) & 42(18) &101(32) \\
%% from crab:/home/matsu/fim21/silLRGel/redf22
\sidehead{Ellipsoid $2:\sqrt{2}:1$, S}
%\sidehead{Ellipsoid $2:\sqrt{2}:1$, $a_{eq}=0.1-1.0\mu$m, $\lambda$=2.2$\mu$m}
 0.4 & 0.10 &   0 & -      & -      & -      & -      & -      \\
 0.6 & 0.16 &  40 & 70(11) & 53( 8) &120(10) & 28( 9) &100(11) \\
 0.8 & 0.23 &  29 & 70(14) & 42( 9) &127(12) & 35(11) & 97(14) \\
 1.0 & 0.30 &  21 & 75(10) & 37( 6) &128(12) & 33(11) & 98(12) \\
\enddata
\tablenotetext{a}{
The letter 'L' stands for the size distribution of $a_{eq}=0.15-1.5\mu$m, 
and the 'S' for $a_{eq}=0.1-1\mu$m, respectively.
The observed values in the K-band ($\lambda=2.2\mu$m) for $p_c^{obs}$, 
$p_l^{obs}$, and $PA^{obs}$ are set as 15\%, 38\%, and 0$\arcdeg$, 
and the ranges of acceptance specified by 
$\delta p_c^{obs}$, $\delta p_l^{obs}$, and $\delta PA^{obs}$ 
as 3\%, 7\%, and 5$\arcdeg$, respectively.
The ranges are larger than the observational errors. }
\end{deluxetable}

\clearpage

\begin{deluxetable}{ccrrrrrr}
\tabletypesize{\scriptsize}
% \rotate
\tablecaption{Models that Explain Observations in the L-band$^a$ \label{tbl-3}}
\tablewidth{0pt}
\tablehead{
\colhead{$R$} & \colhead{$|p_c|^{max}$} & \colhead{$N$} &
\colhead{$\theta_i'$} & \colhead{$\theta_s'$} & \colhead{$\phi_s'$} &
\colhead{$PA'$} & \colhead{$\Theta_{sca}$}
}
\startdata
%% /home/matsu/tmat01/silLRG2/pwr35/redf2_36/15
\sidehead{Oblate 2:1, L}
 0.2 & 0.05 &   0 & -      & -      & -      & -      & -      \\
% 0.3 & 0.07 &  25 & 78( 8) & 50( 7) &112( 9) & 17( 7) & 98( 9) \\
 0.4 & 0.10 &  71 & 78(13) & 48(11) &117(11) & 21(11) &101(13) \\
% 0.5 & 0.12 &  73 & 77(15) & 42(11) &123(16) & 28(15) &101(17) \\
 0.6 & 0.14 &  62 & 77(18) & 40(11) &129(17) & 33(16) &102(19) \\
% 0.7 & 0.17 &  53 & 78(19) & 38(10) &134(19) & 37(18) &105(20) \\
 0.8 & 0.19 &  48 & 76(24) & 39(12) &137(20) & 41(20) &105(25) \\
% 0.9 & 0.22 &  43 & 77(30) & 39(13) &135(26) & 38(25) &105(31) \\
 1.0 & 0.25 &  44 & 76(35) & 42(17) &135(31) & 39(30) &106(36) \\
%
%% /home/matsu/tmat01/silLRG2o3/pwr35/redf2_36
\sidehead{Oblate 3:1, L}
 0.2 & 0.06 &   0 & -      & -      & -      & -      & -      \\
% 0.3 & 0.09 &  60 & 91(14) & 57( 9) &108(10) &  3(12) &106(12) \\
 0.4 & 0.11 &  73 & 87(17) & 51(11) &115(11) & 13(13) &107(14) \\
% 0.5 & 0.13 &  74 & 83(17) & 49(11) &120(12) & 20(14) &107(16) \\
 0.6 & 0.16 &  66 & 84(22) & 44(11) &123(15) & 23(16) &108(21) \\
% 0.7 & 0.18 &  51 & 89(27) & 41(10) &123(20) & 21(21) &110(25) \\
 0.8 & 0.20 &  40 & 94(30) & 38( 9) &122(27) & 17(27) &113(30) \\
% 0.9 & 0.22 &  33 & 96(32) & 39( 8) &119(34) & 13(32) &113(33) \\
 1.0 & 0.24 &  16 &100(29) & 37( 6) &119(39) & 12(36) &115(32) \\
%
%% /home/matsu/tmat01/silLRG/pwr35/redf2_36
\sidehead{Oblate 2:1, S}
 0.4 & 0.06 &   0 & -      & -      & -      & -      & -      \\
 0.6 & 0.09 &   7 &112( 9) & 61( 9) &111( 4) &-16(10) &118( 5) \\
 0.8 & 0.12 &  12 &119(15) & 47( 9) &106( 9) &-18(13) &120(11) \\
 1.0 & 0.15 &  15 &125(15) & 39(10) & 98(17) &-24(15) &121(15) \\
%
%% /home/matsu/tmat01/silLRG2pr/pwr35/redf2_36
\sidehead{Prolate 2:1, L}
 0.4 & 0.05 &   0 & -      & -      & -      & -      & -      \\
 0.6 & 0.07 &  20 & 68( 9) & 57( 9) &110( 7) & 24( 7) & 93( 9) \\
% 0.7 & 0.09 &  53 & 72(13) & 60(13) &108( 9) & 19(11) & 95(12) \\
 0.8 & 0.10 &  75 & 73(16) & 59(13) &109(11) & 18(13) & 97(14) \\
% 0.9 & 0.12 &  81 & 77(17) & 53(15) &106(14) & 14(14) & 95(16) \\
 1.0 & 0.15 &  81 & 83(19) & 48(13) &103(17) &  8(15) & 95(18) \\
%
%% /home/matsu/tmat01/silLRG2p3_36/pwr35/redf2_36
\sidehead{Prolate 3:1, L}
 0.6 & 0.05 &   0 & -      & -      & -      & -      & -      \\
% 0.7 & 0.07 &  11 & 72( 8) & 60( 6) & 85( 6) &  7( 7) & 77( 6) \\
 0.8 & 0.09 &  42 & 80(10) & 60(11) & 82( 8) &  0( 9) & 78( 9) \\
% 0.9 & 0.12 &  74 & 84(11) & 59(13) & 83(11) & -4(10) & 81(11) \\
 1.0 & 0.16 &  56 & 89(12) & 56(14) & 80(16) &-10(13) & 82(15) \\
%
%% /home/matsu/tmat01/silLRGpr/pwr35/redf2_36
\sidehead{Prolate 2:1, S}
 1.0 & 0.07 &   0 & -      & -      & -      & -      & -     \\
%
%% /home/matsu/fim21/silLRG2el-2/pwr35/redf2_36
\sidehead{Ellipsoid $2:\sqrt{2}:1$, L}
 0.2 & 0.04 &   0 & -      & -      & -      & -      & -      \\
 0.4 & 0.08 &  13 & 78(12) & 46( 6) &118(15) & 23(13) &100(13) \\
% 0.5 & 0.10 &  24 & 76(17) & 43(12) &119(14) & 25(13) & 98(17) \\
 0.6 & 0.12 &  39 & 79(16) & 38(11) &118(21) & 23(18) & 98(18) \\
% 0.7 & 0.14 &  35 & 80(19) & 37(12) &118(22) & 22(19) & 98(20) \\
 0.8 & 0.17 &  34 & 75(23) & 38(15) &127(21) & 33(19) & 99(24) \\
% 0.9 & 0.19 &  33 & 76(25) & 38(16) &125(26) & 31(23) & 99(27) \\
 1.0 & 0.21 &  32 & 78(31) & 40(18) &117(37) & 22(32) & 97(34) \\
% 
%% /home/matsu/fim21/silLRGel/redf2_36
\sidehead{Ellipsoid $2:\sqrt{2}:1$, S}
 0.6 & 0.07 &   0 & -      & -      & -      & -      & -      \\
 0.8 & 0.09 &   1 &129( 0) & 51( 0) &101( 0) &-32( 0) &121( 0) \\
 1.0 & 0.11 &   4 &135(13) & 41( 0) & 93(11) &-37(10) &123(10) \\
\enddata
\tablenotetext{a}{
The letter 'L' stands for the size distribution of $a_{eq}=0.15-1.5\mu$m,
and the 'S' for $a_{eq}=0.1-1\mu$m, respectively.
The observed values in the L-band ($\lambda=3.6\mu$m) for $p_c^{obs}$,
$p_l^{obs}$, and $PA^{obs}$ are set as 9.4\%, 57\%, and 0$\arcdeg$,
and the ranges of acceptance specified by 
$\delta p_c^{obs}$, $\delta p_l^{obs}$, and $\delta PA^{obs}$ 
as 3\%, 11\%, and 5$\arcdeg$, respectively.
Ths raqnges are larger than the observational errors. }
\end{deluxetable}

\end{document}